%% file: main.tex
\documentclass[%
nofootinbib,
 amsmath,amssymb,
 aps,
 prd,
 twocolumn,
 superscriptaddress
]{revtex4-1}

\usepackage{dcolumn}% Align table columns on decimal point

\allowdisplaybreaks

\usepackage[dvipdfmx]{graphicx}
\usepackage{latexsym}
\usepackage{amsfonts}
\usepackage{amssymb}
\usepackage{amsmath}
\usepackage{bm}
\usepackage{mathrsfs}
\usepackage[dvipdfm]{hyperref}
\newcommand{\bp}{{\bm{p}}}
\newcommand{\bq}{{\bm{q}}}

\newcommand{\tr}{ {\rm Tr} }
\newcommand{\Lag}{ {\mathscr{L}} }

\newcommand{\M}{ {\mathcal M} }

\newcommand{\ext}{ {\rm ext} }
\newcommand{\exts}{  {\scriptscriptstyle \ext}  }
\newcommand{\vac}{ {\rm vac} }

\newcommand{\as}{ \alpha_s }
\newcommand{\aem}{ \alpha_{\rm em} }
\newcommand{\EM}{ {\rm em} }

\newcommand{\scJ}{ {\scriptstyle  J} }

\newcommand{\swave}{{ \scriptscriptstyle 1S }}
\newcommand{\cc}{{ \scriptscriptstyle {c \bar c} }}
\newcommand{\Res}{{ \rm Res }}
\newcommand{\eb}{{ \epsilon_0}}

\newcommand{\pv}{{_{\scriptscriptstyle  \rm PV}}}
\newcommand{\va}{{_{\scriptscriptstyle \rm VA} }}
\newcommand{\sa}{{_{\scriptscriptstyle \rm SA} }}
\newcommand{\ps}{ {\scriptscriptstyle \rm P} }
\newcommand{\V}{ {\scriptscriptstyle \rm V} }
\newcommand{\s}{ {\scriptscriptstyle \rm S} }
\newcommand{\A}{ {\scriptscriptstyle \rm A} }
\newcommand{\gam}{ {\scriptscriptstyle \gamma} }

\newcommand{\etac}{ {\eta_c} }
\newcommand{\Jp}{ {J/\psi} }

\newcommand{\ph}{ {\rm ph} }
\newcommand{\dir}{ {\rm dir} }

\newcommand{\beq}{\begin{eqnarray}}
\newcommand{\eeq}{\end{eqnarray}}

\def\simge{\mathrel{%
   \rlap{\raise 0.511ex \hbox{$>$}}{\lower 0.511ex \hbox{$\sim$}}}}
\def\simle{\mathrel{
   \rlap{\raise 0.511ex \hbox{$<$}}{\lower 0.511ex \hbox{$\sim$}}}}
\def\bigs{\mathrel{
   \rlap{\raise 0.531ex \hbox{$>$}}{\lower 0.531ex \hbox{$<$}}}}

\usepackage[normalem]{ulem}  % \sout{old text} for strikeout
\usepackage[dvips]{color} % For blue in-text comments and additions

\renewcommand\sout{\bgroup \color{red} \ULdepth=-.5ex \ULset}

\usepackage{slashed}

\begin{document}
%%%%%%%%%%%%%%%%%%%%%%%%%%%%%%%%%%%%%%%%%%%%%%%%%%%%%%%%%%%%%%%%%%%%%%%%%%%%%%%%%%%%

\vspace*{-10mm}
% \begin{flushright}
{\small
\noindent
BNL-107204-2014-JA, RBRC-1102, RIKEN-QHP-172, 

\noindent
KEK-TH-1781, YITP-14-90
}
% \end{flushright}
\vspace{-5mm}

\input{head}

%%%%%%%%%%%%%%%%%%%%%%

%%%%%%%%%%%%%%%%%%%%%%%%%%%%%%%%%%%%%%%%%%%%%%%%%%%%%%%%%%%%%%%%%%%%%%%%%%%%%%

\section{Introduction}

It has been known for quite some time that external magnetic fields 
strongly interacting with charged fermions give rise to intriguing dynamics in vacuum, 
including not only nonlinear dynamics of photons within QED \cite{QED, HI} 
but also an interplay with QCD. 
Motivated by formation of strong electromagnetic fields 
in neutron stars/magnetars \cite{TD, HL06} and ultrarelativistic heavy-ion collisions
at the Relativistic Heavy Ion Collider (RHIC) and the Large Hadron Collider (LHC) 
\cite{SBS, KMW, Bestimates, Itakura_PIF}, 
a number of lattice QCD simulations and analytic calculations have shown that 
strong magnetic fields modify properties of QCD vacuum 
such as quark condensates \cite{SBS, IS, BCat, GMS, RG, ChPT, SD1, qbarq_lat, BMW_qbarq, sus_lat, SD2} 
and gluon condensates \cite{BMW_GG, Ozaki}, 
and consequently modify even phase structures \cite{qbarq_lat, BMW_GG, PT, FRG, CY} 
and hadron properties \cite{SBS, IS, GOR, Chern, HY, rho_lat, QP_lat, rho_NJL, MT, Mac, AS, Machado, Letter, HQP}. 
One of the remarkable findings is a discrepancy between %charged $\rho$
meson spectra obtained from a hadronic effective model calculation
and a lattice QCD simulation in strong magnetic fields \cite{HY, rho_lat}.
A lesson learned there might be the importance of studying bound-state properties
on the basis of elementary degrees of freedom in the underlying theory,
when the  magnitudes of external fields approach and go beyond the typical scales of
the theory. Especially in QCD, this is important not only
because an internal structure of a bound state is changed
but also because properties of QCD vacuum are changed.
Therefore, these observations pose a fundamental question in QCD, i.e.,
how changes of QCD vacuum are reflected in meson spectra in external fields.

In a recent paper \cite{Letter}, we investigated a field theoretical approach to this issue, 
and proposed a general framework of the QCD sum rules (QCDSR) applied to meson spectroscopy in
external magnetic fields. This paper is supposed to be
a detailed account of our framework providing a semi-analytic
method to elucidate the relation between properties of QCD vacuum
and meson spectra in external magnetic fields. We also extend our detailed analysis to include the results for a transversely polarized $J/\psi$ with respect to the direction of the external field.   Historically, the
QCD sum rule was developed soon after the discovery of $J/\psi$
and first applied to heavy-quark systems \cite{Nov77,SVZ791,SVZ792}. 
The QCD sum rules remarkably predicted the small mass splitting between $\eta_c$
and $J/\psi$, which was subsequently confirmed by experiments
\cite{etac}. The resolution in the results from the QCDSR was as high as  
the order of the mass splitting of less than 100 MeV. This was
achieved by taking into account effects of a gluon condensate as
well as a perturbative piece in current correlators on the basis
of the operator product expansion (OPE) \cite{Wil69}. As shown in the
seminal papers \cite{SVZ792, RRY80, RRY81}, the QCD sum rules allow for manifestly
incorporating nonperturbative effects of QCD vacuum through
expectation values of operators in the OPE \cite{RRYrev, Shifman, Narison}, which can be performed 
on an order-by-order basis with respect to mass dimensions of the
operators and thus in a systematic manner. This structure in QCD
sum rules indicates that we could investigate how changes in
properties of QCD vacuum in the OPE manifest %incorporated through the OPE manifest
themselves in hadron properties in external environments. Indeed,
the QCD sum rules at finite temperature/density \cite{BS86, Furnstahl-Hatsuda-Lee90, Coh} 
have been applied to heavy quarkonia 
\cite{KKLW, Hay, ML, ML08, LM09, SLM, ML10, GMO, ML12, SGMO, Lee:2013dca}, 
light mesons \cite{HatsudaLee, light}, and heavy-light mesons \cite{HL} in the last two decades, 
and very recently those in strong external magnetic fields \cite{Machado,Letter}.

In particular, the heavy quarkonia have been investigated by
various methods as well as the QCD sum rules, since dissociation of
quarkonia in hot media with liberated color degrees of freedom,
the so-called ``$J/\psi$ suppression'', was proposed as a
signature of the quark-gluon plasma created in early times after
the ultrarelativistic heavy-ion collisions \cite{MS,Hashi}. Since then,
not only the hot medium effects but also other effects, such as
the cold nuclear matter effect, regeneration of melted charm quark pair, etc, 
have been examined in theoretical and experimental
studies (see, e.g., Ref.~\cite{CCreviews} for reviews). While the strong
electric and magnetic fields rapidly decay in the early-time
dynamics \cite{Bestimates}, they could also act on
heavy quarkonia and give rise to measurable effects \cite{MT,AS}
because estimates on heavy-quarkonium formation time indicate a
prompt formation in early times \cite{formation} where the strong
electromagnetic fields still persist with large magnitudes. 
Estimates on formation times given in Ref.~\cite{formation} have shown %suggest 
that a significant fraction of charmonium and/or bottomonium produced in the heavy-ion collisions 
will be formed faster than 0.2 fm/c where the strength is still 
in the range from $0.1 m_\pi^2$ to $m_\pi^2$ at RHIC energies.

Motivated by these theoretical and phenomenological aspects, we
investigate $\eta_c$ and $J/\psi$ in the presence of external
magnetic fields in detail by using hadronic effective theories and
the QCD sum rules. While we focus on charmonia in this paper, the
same methods can be applied to corresponding bottomonium states.
We first give a systematic analysis of mixing patterns among
charmonia in terms of a hadronic effective theory. We will find
that only a mixing between a $\eta_c$ and a longitudinal $J/\psi$
is possible when charmonia are at rest, where a longitudinal
$J/\psi$ is meant for a state with a vanishing spin component with
respect to the direction of an external magnetic field. A level
repulsion due to this mixing effect is consistent with results in
a preceding study in terms of a potential model \cite{AS}. Bearing
this in mind, we will switch to the QCD sum rules to investigate
charmonia on the basis of the fundamental degrees of freedom, and
elaborate the hadronic spectral density ansatz called the
phenomenological side to consistently take into account the mixing
effects \cite{Letter}. We will show how to distinguish nonperturbative mass
modifications from a level repulsion from the mixing effect that
can be described on the hadronic basis. We note that our
treatment of the mixing effects can be applied to general analyses
on meson spectra in terms of correlation functions, and should be
applied to a very recent QCDSR analysis on $B$ mesons in strong
magnetic fields \cite{Machado} since the $B$ mesons are mixed with
$B^\ast$ mesons. Our work demonstrates how to implement mixing
effects in the QCDSR method, in particular for the heavy quark
systems where both the OPE and the phenomenological side are well
under control, and thus provides a general guideline to include
mixing effects in approaches based on correlation functions.

Operator product expansion is then implemented up to dimension-4
operators in which we have external magnetic fields $\langle
F_{\alpha\beta} F_{\mu\nu} \rangle$ as operator expectation
values, in addition to a scalar gluon condensate $\langle
G^a_{\mu\nu} G^{a \, \mu\nu} \rangle$ common to the OPE in
the ordinary vacuum. It is noteworthy that the dominant effects in
finite temperature/density comes into the OPE only through
dimension-4 gluon condensates that are related to the
energy momentum tensor of which the matrix elements are well estimated
both at finite temperature from lattice QCD
\cite{ML,ML08,ML10,GMO,SGMO} and at normal nuclear matter density
from measurements in deep inelastic scatterings \cite{KKLW}.
Recently, it has also been shown that the strength of the charmonium wave function at
the origin obtained from the QCDSR follows precisely that obtained from 
solving the
Schr\"odinger equation with a finite temperature free energy
potential extracted from lattice QCD \cite{Lee:2013dca}. In cases of
external magnetic fields, it would become necessary to resum
all-order terms with respect to dimensions of external fields
$\langle FFF \cdots \rangle$ when the magnitude of a magnetic field
goes beyond a separation scale in the OPE as recently performed
for a vector current correlator \cite{HI}.

Based on these elaborate treatments both on the phenomenological and
the OPE sides, results of mass modification from 
the QCD sum rules are found to be consistent with those from the mixing
effects with some slight discrepancies. 
% On the basis of these results from a QCD-based framework, 
We will then argue that the dispersion relation in the QCD sum rules
is saturated by the mixing-induced terms, and identify the mixing
effect as the dominant origin of mass shifts of static $\eta_c$
and the longitudinal $J/\psi$. Then, we will examine effects of a perturbative
heavy-quark loop as a subdominant origin of mass shifts in those states and 
of a transverse $J/\psi$ that is not involved in the mixing pattern.

This paper is organized as follows. We first examine possible
mixing patterns in terms of a hadronic effective theory in
Sec.~\ref{sec:hadron}, followed by analyses with the use of the QCD
sum rules in the subsequent sections. After a brief description of
the QCD sum rule for heavy quarkonia in
Sec.~\ref{sec:description}, we elaborate on the phenomenological side
bearing the mixing pattern in mind in Sec.~\ref{sec:phen}, and
implement the OPE in the presence of an external magnetic field in
Sec.~\ref{sec:OPE}. Combining these ingredients, mass spectra of
static $\eta_c$, longitudinal $J/\psi$ and transverse $J/\psi$
from the QCD sum rule are obtained as shown in
Sec.~\ref{sec:QCDSR} with discussion about the role of
mixing-induced terms on the phenomenological side and the origins of
residual mass shift other than the mixing effect in Secs. 
\ref{sec:Bterms} and \ref{sec:SE}, respectively. Section
\ref{sec:summary} is devoted to the summary. In the Appendix, we provide
a list of the Wilson coefficients and some details of the
calculations on the hadronic basis including calculation of the
coupling constants in the mixing effects by using Bethe-Salpeter
amplitudes of charmonia.

%%%%%%%%%%%%%%%%%%%%%%%%%%%%%%%%%%%%%%%%%%%%%%%%%%%%%%%%%%%%%%%
%%%%%%%%%%%%%%%%%%%%%%%%%%%%%%%%%%%%%%%%%%%%%%%%%%%%%%%%%%%%%%%

%#! latex main.tex

\section{Mixing effects in external magnetic fields}

\label{sec:hadron}

\begin{table}[t]
\begin{center}
{\renewcommand\arraystretch{1.7}
 \begin{tabular}[c]{ @{\vrule width 1.3pt } c @{\; \vrule width 1.3pt  }
c|c|c|c @{\vrule width 1.2pt} } \hline \noalign{\hrule height 1.2pt}
                & $\eta_{c}$ & $J/\psi$ & $\chi_{c0}$ & $\chi_{c1}$ \\ \hline \noalign{\hrule height 1.3pt}
%  &&&& \\ \hline
\;\;\;\;\; $\eta_{c}$ \;\;\;\; & {\LARGE \bf ---}  & \; $ P-V_\parallel $ \;
& \hspace{0.5cm}  -- \hspace{0.5cm}  & -- \\ \hline
$J/\psi$ & \;\, $ V_\parallel - P $ \;\, & {\LARGE \bf ---}  & -- &
\hspace{0.5cm}  -- \hspace{0.5cm}  \\ \hline
\, $\chi_{c0}$ & -- & -- &  {\LARGE \bf ---}  &  $S-A_\parallel $   \\ \hline
\, $\chi_{c1}$ & -- & \hspace{0.5cm}  -- \hspace{0.5cm} &
 $A_\parallel-S $  &  {\LARGE \bf ---}   \\ \hline
\noalign{\hrule height 1.3pt}
  \end{tabular}
}
  \end{center}
\caption{Summary of possible mixing patterns for the quarkonia at
rest in external magnetic fields. Subscripts denote a longitudinal 
components of the vector and the axial-vector fields introduced in the text. 
% Effective interaction vertices are shown in Eqs.~(\ref{eq:Eq_P})--(\ref{eq:Eq_A}).
}
\label{tb:mixing}
\end{table}

% \begin{figure}[t]
% \begin{center}
% \includegraphics[width=\hsize]{../figs/SE.eps}
% \end{center}
% \caption{Mixing effects in an external magnetic field.
% Double solid and wavy lines show a charmonium %in external lines
% and its mixing partner appearing in the intermediate state, respectively.
% The second diagram on the right--hand side shows contribution of a heavy--quark loop
% shown in Fig.~\ref{fig:SE}.}
% \label{fig:quad}
% \end{figure}

We first examine effects of external magnetic fields on charmonia
%described by 
in terms of mesonic degrees of freedom.
One should notice that even neutral mesons can be affected by
external magnetic fields through effective interaction vertices,
and that any state can appear in the intermediate states as long as quantum numbers are matched.
We thus investigate what mixing patterns are possible in external magnetic fields
among the low-lying charmonia, $\eta_c$, $J/\psi$, $\chi_{c0}$, and $\chi_{c1}$. 
This can be systematically discussed in terms of a hadronic effective Lagrangian
constrained by symmetries of the system as shown below. 
% The mixing patterns we find are summarized in Table~\ref{tb:mixing}.

We investigate mixing effects among the pseudoscalar $(\eta_c)$,
vector $(J/\psi)$, scalar $(\chi_{c0})$ and axial-vector
$(\chi_{c1})$ quarkonia by a hadronic effective Lagrangian
approach. An effective Lagrangian includes all the relevant
three-point vertices among two static quarkonia and a photon (external magnetic field), 
\begin{eqnarray}
\Lag = \Lag_{ {\rm kin} + M } + \Lag_{\gam \pv} + \Lag_{\gam \va} + \Lag_{\gam \sa}
\label{eq:Lag}
\end{eqnarray}
where the kinetic and mass terms are as usual given by
\begin{eqnarray}
\Lag_{ {\rm kin} + M } &=& - \frac{1}{2} \partial_{\mu} P \partial^{\mu} P + \frac{1}{2} m_{\ps}^{2} P^{2}
\nonumber
\\
&& %\hspace{0.3cm}
- \frac{1}{2} \partial_{\mu} V_{\nu} \partial^{\mu} V^{\nu} + \frac{1}{2} m_{\V}^{2} V^{2}
\nonumber
\\
% && %\hspace{0.3cm}
%  - \frac{1}{2} \partial_{\mu} S \partial^{\mu} S + \frac{1}{2} M_{\s}^{2} S^{2}
% \nonumber
% \\
% && %\hspace{0.3cm}
% - \frac{1}{2} \partial_{\mu} A_{\nu} \partial^{\mu} A^{\nu} + \frac{1}{2} M_{\A}^{2} A^{2}
&&
+ (P \rightarrow S) + (V \rightarrow A)
\ .
\end{eqnarray}
The pseudoscalar and the vector fields are denoted by $P$ and $V^\mu$, respectively,
and those terms for the scalar field $(S)$ and the axial-vector field $(A^\mu)$ are
given by the replacements indicated in the last line.
Possible interaction vertices among those fields, and thus mixing patterns, are informed from
the Lorentz invariance and the parity and charge-conjugation symmetries. 
The vertices relevant for interactions among static charmonia are found to be 
\begin{eqnarray}
\Lag_{\gam \pv} &=&
\frac{ g_{\pv} }{ m_0 } e \tilde{F}_{\mu \nu} (\partial^{\mu} P) V^{\nu}
\label{eq:L_pv} \ ,
\\
\Lag_{\gam \va}
&=& i g_{\va} e \tilde{F}_{\mu \nu} V^{\mu} A^{\nu}
\label{eq:L_va} \ ,
\\
\Lag_{\gam \sa}
&=& \frac{ g_{\sa} }{ m_1 } e \tilde{F}_{\mu \nu} (\partial^{\mu} S) A^{\nu}
\label{eq:L_sa}
\ ,
\end{eqnarray}
with $m_0=(m_\ps+m_\V)/2$, $m_1=(m_\s+m_\A)/2$,
and dimensionless effective coupling constants $g_\pv$, $g_\va$, and $g_\sa$.
These vertices are responsible for, e.g., radiative decay modes of quarkonia
such as $\Jp \rightarrow \eta_c + \gamma$.

Note that interaction vertices proportional to the field strength
tensor $F^{\mu\nu}$, such as $\Lag_{\gam {\scriptscriptstyle \rm
VS}} \propto F_{\mu\nu} (\partial^\mu S) V^\nu$, do not play a
role when addressing mixing effects among the static quarkonia
in external magnetic fields, and are not shown above. 
Since the field strength tensor $F^{\mu\nu}$ has finite elements 
only in the spatial components in case of an external magnetic field, 
it inevitably picks up vanishing spatial momenta of quarkonia 
when contracted with the derivatives, i.e., $F_{\mu\nu} \partial^\nu = 0$,  
and does not get involved in any mixing effect
addressed here. Therefore, the interaction vertices should be
proportional to the dual field strength tensor $\tilde F^{\mu\nu}$
as those in Eqs.~(\ref{eq:L_pv})-(\ref{eq:L_sa}). Note, however,
that a coupling between the vector and axial-vector mesons
(\ref{eq:L_va}) does not introduce any physical interaction,
because the nonvanishing component of $\tilde F^{\mu\nu}$ picks
up an unphysical temporal component of either the vector or
axial-vector field.

Following from the discussions above, 
we eventually found that only two mixing patterns, one between $\eta_c$ and $J/\psi$
and the other between $\chi_{c0}$ and $\chi_{c1}$, are possible when they are at rest in external magnetic fields. 
These results are summarized in Table~\ref{tb:mixing}. 
Since neither $\chi_{c0}$ nor $\chi_{c1}$ is mixed with $\eta_c$ and $J/\psi$,
we shall focus on $\eta_c$ and $J/\psi$ in the present work,
and calculate the mass eigenstates in the presence of the mixing effects
by solving equations of motion which follow from the effective Lagrangian \eqref{eq:Lag} as
\begin{eqnarray}
&& \hspace{-0.5cm} P \!:
( \partial^{2} + m_{\ps}^{2} ) P
- \frac{g_{\pv} }{m_{0}} e \tilde{F}_{\alpha \beta} \partial^{\alpha} V^{\beta} = 0
\label{eq:Eq_P}
,
\\
&& \hspace{-0.5cm} V  \! :
( \partial^{2} + m_{\V}^{2}) V_{\mu} + \frac{g_{\pv} }{ m_{0} } e \tilde{F}_{\alpha \mu} \partial^{\alpha} P
 = 0
\label{eq:Eq_V}
.
% \\
% && \hspace{-0.5cm} S \!: \
% ( \partial^{2} + M_{\s}^{2} ) S
% - \frac{g_{\sa} }{M_{\s}} \tilde{F}_{\alpha \beta} \partial^{\alpha} A^{\beta} = 0
% \label{eq:Eq_S}
% ,
% \\
% && \hspace{-0.5cm} A \!: \
% ( \partial^{2} + M_{\A}^{2} ) A_{\mu} + \frac{g_{\sa} }{ M_{\s} } \tilde{F}_{\alpha \mu} \partial^{\alpha} S
% +  ig_{\va} \tilde{F}_{\alpha \mu } V^{\alpha} = 0
% \label{eq:Eq_A}
% .
\end{eqnarray}
To show the mixing patterns more clearly, we hereafter assume that
an external magnetic field is oriented in the positive $z$-direction, %as in the calculation of the QCDSR, 
where the dual field strength tensor has only two nonzero components
% $\tilde{F}^{30} = - \tilde{F}^{03} =  B$.
$\tilde{F}_{03} = - \tilde{F}_{30} =  B$.
In this configuration, the vector field reads %has the four components as
$V^\mu = (V_0, \bm V_\perp, V_\parallel )$
where $V_0$, $\bm V_\perp $, and $V_\parallel $
denote the temporal, two transverse and one longitudinal modes
with respect to the external magnetic field, respectively.

\begin{figure}[t]
 \centering
 \includegraphics[width=\columnwidth]{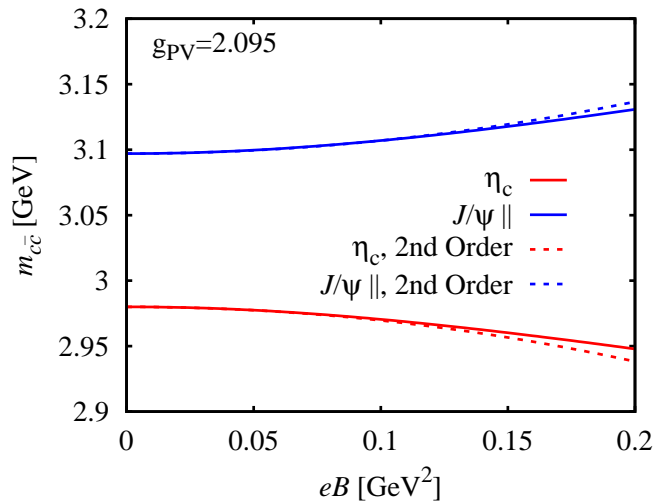}
\caption{
Mixing effects between static $\eta_{c}$ and the longitudinal $J/\psi$.
Solid (dotted) lines show a level repulsion from the mixing effects
in all-order (second-order) with respect to $eB$.
}
\label{fig:Qshift}
\end{figure}

With a vanishing spatial momentum $q^\mu = (\omega, 0 , 0, 0)$,
the equations of motion (\ref{eq:Eq_P}) and (\ref{eq:Eq_V})
result in a $2\times 2$ matrix form:
\beq
\left(
\begin{array}{cc}
- \omega^{2} +  m_{\ps}^{2} & - i \frac{g_{\pv}}{m_{0}} \omega e B \\
i \frac{g_{\pv}}{m_{0}} \omega e B & -\omega^{2} +  m_{\V}^{2} \\
\end{array}
\right)
\left(
\begin{array}{c}
P \\
V_{\parallel} \\
\end{array}
\right)
\label{Matrix_eq_etac_Jpsi}
=0.
\eeq
We notice that a mixing is held only between $\eta_c$ and the longitudinal $J/\psi$, 
and that the transverse $J/\psi$ is not mixed with $\eta_c$, 
as summarized in Table~\ref{tb:mixing}. 
Following from the equations of motion (\ref{Matrix_eq_etac_Jpsi}),
we obtain the physical mass eigenvalues in the presence of the mixing effect as
\begin{equation}
%\hspace{-0.5cm}
m_{\Jp,\etac}^{2}=\frac{1}{2} \bigg( M_+^2+\frac{\gamma^2}{m_{0}^{2}}
\pm \sqrt{M_-^4+\frac{2\gamma^2 M_+^2}{m_0^2}+ \frac{\gamma^4}{m_0^4}} \bigg), 
\label{eq:EFT}
\end{equation}
where $M_+^2=m_\ps^2+m_\V^2, M_-^2=m_\V^2-m_\ps^2$ and
$\gamma=g_{\pv} e B$. Expanding Eq.~\eqref{eq:EFT} up to the
second order in $\gamma$ and the leading order in
$\frac{1}{2}(m_\V-m_\ps)/m_{0} $, %$\sim 0.02$
we find
\beq
m_{\Jp,\etac}^{2} &=&
m_{\V,\ps}^{2} \pm \frac{\gamma^2}{M_-^2}
,
\label{eq:Jpsi_2nd}
%,
%\\
%M_\etac^{2}
%&=& M_{\ps}^{2}  +  \frac{ g_{\pv}^2 (e B)^{2} }{ M_{\ps}^{2} - M_{\V}^{2} }
%.
%\label{eq:eta_2nd}
\eeq
with eigenvectors given by
\beq
% | \etac \rangle_B \ \, &=&  \bigg(1-\frac{1}{2} \gamma^2 \Delta^2  \bigg)
% | \etac \rangle - \gamma \Delta
% | \Jp \rangle   , \nonumber \\
% | \Jp \rangle_B &=& \gamma \Delta
% | \etac \rangle +  \bigg(1-\frac{1}{2} \gamma^2 \Delta^2 \bigg)
% | \Jp \rangle
| \etac )_{\scriptscriptstyle B} \ \, &=&  \bigg(1-\frac{1}{2} \frac{\gamma^2}{ M_-^4}  \bigg)
| P ) - i \frac{\gamma}{M_-^2}
| V )   , \nonumber \\
| \Jp )_{\scriptscriptstyle B} &=& -i \frac{\gamma}{M_-^2}
| P ) +  \bigg(1-\frac{1}{2} \frac{\gamma^2}{M_-^4} \bigg)
| V ) .
\label{eq:wave}
\eeq
We show plots of the mass shifts in the presence of the mixing effects in Fig.~\ref{fig:Qshift}.
The coupling constant $g_\pv=2.095$ is here obtained by
fitting the radiative decay widths measured in experiments. 
See Appendix~\ref{sec:coupling} for details.
We also show that the effective coupling between $\eta_c$ and $J/\psi$
can be obtained from the mixing amplitudes computed by utilizing
Bethe-Salpeter amplitudes \cite{BSamp},
and that the coupling strength agrees well with the one from the fitting method
(see Appendix.~\ref{sec:coupling_FS}).
In Fig.~\ref{fig:Qshift}, we find that the mass of $\eta_{c}$ decreases as $eB$ increases,
while the mass of the longitudinal mode of $J/\psi$ (denoted by $J/\psi\, ||$)increases,
indicating a level repulsion between these mass eigenstates in an external magnetic field.
These behaviors are consistent with what was obtained in the potential-model approach \cite{AS},
in which the authors found a level repulsion between $\eta_{c}$ and the longitudinal $J/\psi$
by solving Schr\"odinger equations in the presence of an external magnetic field.

The mixing effect found above, however, does not exhaust possible
effects of external magnetic fields on charmonia. Since the
Lagrangian (\ref{eq:Lag}) contains only the minimal couplings to 
external magnetic fields, %leading-order vertices, 
further mass shifts could be caused by magnetic fields acting on the loops %the charmed meson loops such as a $D\bar{D}$ loop 
and/or interactions among charmonia and more than two photons
(magnetic fields) as higher-order corrections to the effective vertex (\ref{eq:L_pv}). 
As for the loops effects, there could be fermion loops with light nucleons (nucleon-antinucleon loop) 
or with charmed baryons, and boson loops with light or charmed mesons. 
Among those, the loops with light hadrons are highly suppressed due to the OZI rule. 
The only relevant loop effects are those from charmed mesons such as the $\bar D D$ loops, so that 
we will examine effects of the loop contribution composed 
of charm quarks by the potential nonrelativistic QCD (pNRQCD) approach in Sec.~\ref{sec:SE}. 
To investigate effects of those residual
interactions as well as the mixing effect, we will in the next
section switch to the QCDSR method based on the fundamental degrees of
freedom.

%%%%%%%%%%%%%%%%%%%%%%%%%%%%%%%%%%%%%%%%%%%%%%%%%%%%%%%%%%%%%%%
%%%%%%%%%%%%%%%%%%%%%%%%%%%%%%%%%%%%%%%%%%%%%%%%%%%%%%%%%%%%%%%

%#! latex main.tex

\section{Generalities in QCD sum rule for heavy quarkonia}

\label{sec:description}

% \subsection{Dispersion relations}

We provide a concise description of the QCD sum rule in
application to quarkonium spectroscopy \cite{SVZ792, RRY80, RRY81, RRYrev} 
used to investigate mass spectra of $\eta_c$ and $J/\psi$ in the present paper. Those
charmonium states are respectively created by heavy-quark
currents, 
\begin{eqnarray}
j^P &=& i \bar c \gamma^5 c
\label{eq:jP}
\\
j^V_\mu &=& \bar c \gamma_\mu c
\label{eq:jV}
% \\
% j^S &=& \bar c c
% \label{eq:jS}
% \\
% j^A_\mu &=&
% - \frac{1}{q^2} P_{\mu\nu} \, \bar c \gamma^5 \gamma^\nu  c
% (q_\mu q_\nu/ q^2 - g_{\mu\nu} ) \, \bar c \gamma^5 \gamma^\nu  c
% \label{eq:jA}
% \ \ ,
\end{eqnarray}
where superscripts $P$ and $V$ denote pseudoscalar and vector currents, respectively. 
While one can construct a sum rule for the each channel, 
the following descriptions are common to all of these channels.
% The axial--vector current is here projected in the transverse component,
% and thus that an anomaly--free conserved component is examined below.

Since we investigate charmonia created by the currents (\ref{eq:jP}) and (\ref{eq:jV}),
we should closely look at intermediate states in a current correlator
\begin{eqnarray}
\Pi^\scJ (q) = i \int \!\! d^4\!x \, e^{iq\cdot x} \langle 0 \vert T[ J(x) J(0) ] \vert 0 \rangle
\label{eq:corr}
\ \ ,
\end{eqnarray}
where subscripts $J$ denotes a channel and
the Lorentz indices in the vector current are suppressed for simplicity.
While an imaginary part of the correlator is related to charmonium spectra,
computation of this quantity is by no means easily attainable
for an external momentum in the hadron mass scale, %, $q^2 \sim m_{\rm hadron}^2$,
where the system is governed by nonperturbative effects of QCD in the strong-coupling regime.
On the other hand, the asymptotic freedom in QCD allows for
a series representation by operator product expansion (OPE) \cite{Wil69}
with an external hard momentum $Q^2 = -q^2 \gg \Lambda_{\rm QCD}^2$ as
\begin{eqnarray}
\Pi^\scJ (Q^2)  = C_{\rm p}^\scJ \cdot {\bm 1}
+ \sum_d C^{\scJ (d)} (Q^2) \cdot \langle {\mathcal O }^{(d)} \rangle
\label{eq:OPE0}
\ \ ,
\end{eqnarray}
where a summation index $d$ corresponds to the mass dimension of operators $\mathcal O$.
The first term $ C_{\rm p}^\scJ $ being proportional to unit operator
contains not only the leading-order diagram, i.e., the bare polarization diagram,
but also perturbative corrections with respect to a small value of the QCD coupling constant $\alpha_s(Q^2) \ll 1  $.
The subsequent terms contain nonperturbative corrections, in which
the Wilson coefficients $C^{\scJ (d)} (Q^2)$ account for the hard-scale dynamics
on the basis of a perturbative expansion % with respect to a small QCD coupling constant,
while expectation values of the operators $\langle {\mathcal O }^{(d)} \rangle$
incorporate the soft-scale dynamics \cite{Wil69}.
When quarks and gluons carry soft momenta in the intermediate states in the correlator,
the expectation values of the operators such as the quark condensates $\langle \bar q q \rangle$ and
the gluon condensates $\langle G^a_{\mu\nu} G^{a \, \mu\nu} \rangle$ are necessary for
taking into account nonperturbative interactions with the QCD vacuum \cite{SVZ791,SVZ792}.

The OPE works efficiently when there is a definite separation scale,
which usually resorts to an external hard momentum $Q^2$.
The Wilson coefficient for a dimension $d$ operator
behaves as a negative-power factor $C^{\scJ (d)} \sim (Q^2)^{-d/2} $,
%\changed{up to logarithmic corrections from anomalous dimensions},
and thus contributions of the higher dimensional operators containing nonperturbative corrections
are suppressed by $(Q^2)^{-d/2}$ as the momentum scale goes to the deep Euclidean region, $Q^2 \rightarrow \infty$,
leaving perturbative corrections in the first term in Eq.~(\ref{eq:OPE0}).
In case of a heavy-quark system, it was argued
that the Wilson coefficient scales as $(4 m^2 + Q^2 )^{-d/2}$
\cite{SLM, LPKS}. %(see an argument in Appendix~\ref{sec:counting}):
The OPE is reliable even for a small value of $Q^2 $ since any
positive $Q^2$ in the complex $Q^2$-plane is distant from
singularities originated from physical degrees of freedom, i.e.,
poles and thresholds of continua, owing to the large value of
heavy-quark mass $m$. As long as expectation values of
dimension-$d$ operators are much smaller than the separation scale
$(4 m^2 + Q^2 )^{-d/2}$, one could plausibly perform the OPE. This
is the case for the OPE in the presence of external magnetic fields
expected for the early stage in relativistic heavy ion collisions.
Up to the Large Hadron Collider energies,  the magnetic field
$\vert e \bm B \vert \alt 10 m_\pi^2$ can be induced by colliding nuclei, 
\cite{Bestimates} thus it satisfies a condition
$\vert e \bm B \vert \ll 4m^2 + Q^2 $. We will implement the OPE in
Secs.~\ref{sec:OPEvac} and \ref{sec:OPEext}.

Once the series representation by OPE \eqref{eq:OPE0} is obtained, it
can be related to the spectral density, namely the imaginary part of the
correlator in the physical region $(q^2 > 0)$, through a dispersion
relation
\begin{eqnarray}
\tilde \Pi^\scJ ( Q^2 )  = \frac{1}{\pi} \int_0^\infty
\frac{ \, {\rm Im} \, \tilde \Pi^\scJ (s) \, }{ s + Q^2 } \, ds
\ + \ ({\rm subtraction})
\label{eq:disp0}
.
\end{eqnarray}
We have introduced a dimensionless current correlator $\tilde \Pi^\scJ (Q^2)$ normalized as follows.
The dispersion relation (\ref{eq:disp0}) is satisfied individually with respect to three polarization modes
in the vector channel, so that we will investigate spin-projected scalar correlators
$\tilde \Pi^V = (\epsilon_\mu \Pi^{\V\mu\nu} \epsilon_\nu) / q^2$
specified by polarization vectors $\epsilon^\mu$ as shown in Sec.~\ref{sec:OPEext}.
We will find a mass splitting among spin polarization states in external magnetic fields.
As for the pseudoscalar channel,
we have a dimensionless correlator $\tilde \Pi^\ps = \Pi^\ps/q^2$.

One would be still skeptical to the applicability of the dispersion
relation (\ref{eq:disp0}) to mass spectroscopy of bound states,
since the series representation by the OPE is related only to an
integrated spectral density which includes contributions from not
only all the poles but also continua as a mixture. However, note
that the integrand in Eq.~(\ref{eq:disp0}) is weighted around the
lower boundary of the integral region for a positive value of
$Q^2$, and higher energy contribution to the integrand is suppressed as
the integral variable $s$ goes to
infinity. This trend becomes stronger if the denominator has a
higher power, implying that the integral is eventually dominated by
the contribution from the lowest bound state for a sufficiently large
power. Therefore, we shall take derivatives on both sides of
Eq.~(\ref{eq:disp0}) to suppress the higher energy contribution other
than the lowest pole. Putting the moments of the left-hand side to be
\begin{eqnarray}
M_n^J (Q^2) =  \frac{1}{n!} \left( - \frac{d \ }{dQ^2 } \right)^n \!\! \tilde \Pi^\scJ (Q^2)
\label{eq:Mn}
\ \ ,
\end{eqnarray}
% \begin{eqnarray}
% \frac{1}{n!} \left( - \frac{d \ }{dQ^2 } \right)^n
% \int_0^\infty \frac{ \, {\rm Im} \, \tilde \Pi(s) \, }{ s + Q^2 } \, ds
% = \int_0^\infty \frac{ \, {\rm Im} \, \tilde \Pi(s) \, }{ ( \, s + Q^2 \, )^{n+1}  } \, ds
% \nonumber
% \ \ ,
% \end{eqnarray}
we find {\it the moment sum rule} as %by differentiating the both side as
\begin{eqnarray}
M_n^J(Q^2) =
\frac{1}{\pi} \int_0^\infty \frac{ \, {\rm Im} \, \tilde \Pi^\scJ (s) \, }{ ( \, s + Q^2 \, )^{n+1}  } \, ds
\label{eq:MSR}
\ \ .
\end{eqnarray}
The moment sum rule (\ref{eq:MSR}) was invoked to calculate
charmonium masses, in which the integral in Eq.~(\ref{eq:MSR}) was
carefully examined and was indeed found to be dominated by the
lowest pole contribution as the number of derivatives $n$ becomes
large \cite{SVZ792,RRY81}. Since the Wilson coefficient for a
dimension-$d$ operator in the OPE,
scaling as $(Q^2)^{-d/2}$ or $(4m^2+Q^2)^{-d/2}$, % as mentioned below Eq.~(\ref{}),
has stronger dependence on $Q^2$ than the lower dimension terms,
we notice that contributions from higher-dimension operators,
and thus nonperturbative effects, are enhanced as the number of derivatives becomes larger.
These scaling behaviors with respect to $Q^2$ and $n$ are naturally expected,
because the dominant lowest pole contribution at large $n$ is attributed to nonperturbative effects
while a smeared continuum is described on the basis of a perturbative picture.

The moments of the Wilson coefficients (\ref{eq:Mn}) were obtained
first in a series of seminal papers \cite{SVZ791,SVZ792}, followed
by intensive calculations \cite{RRY80,RRY81,NR82,NR83}. The first
attempt was made at $Q^2=0$ on the basis of an argument that a
convergence of the OPE is, even with vanishing $Q^2$, supported by
a large value of charm quark mass \cite{SVZ791,SVZ792}. However,
it was shown that a better convergence is achieved by taking a
finite momentum square $Q^2 > 0$ \cite{RRY80,RRY81}, and further
that contributions from the higher-dimension operators at a large
value of $n$ can be suppressed only when $Q^2 $ is finite
\cite{RRY84}. Therefore, the momentum square $Q^2$ is preferred to
be taken large. However, if the momentum square $Q^2$ is taken to
be arbitrarily large at a fixed $n$, it spoils the separation of
the lowest pole contribution in Eq.~(\ref{eq:MSR}) because the
integrand is equally suppressed over the whole integral region.
This separation would be restored, if we take a larger $n$ as we
take a large value of $Q^2$ so that a steeper behavior of the
denominator puts a weight on the lowest pole contribution, whereas
convergence of the OPE again becomes weaker for a large $n$ due to
picking up strong $Q^2$-dependence of the Wilson coefficients for
the higher dimension operators in Eq.~(\ref{eq:Mn}). Therefore,
one has to manage to adjust $Q^2$ and $n$ so that the convergence
of the OPE and the separation of the lowest pole contribution are
compatible to each other.

This point would become rather clear if one takes simultaneous limits $Q^2 \rightarrow \infty$ and $n \rightarrow \infty$
while maintaining a constant ratio $ M ^2 := Q^2/n$.
Following conventions in Ref.~\cite{Bertl}, 
we define the limiting form of the moments (\ref{eq:Mn}) as
\begin{eqnarray}
\M_{\text{OPE}}^J (M^2) = 
\lim_{ \substack{Q^2, n \rightarrow \infty \\ Q^2/n = M^2 } }  \pi \, (Q^2)^{n+1} M_n^J (Q^2)
\label{eq:M0}
\ \ ,
\end{eqnarray}
% By taking the same limits on the right--hand side in Eq.~(\ref{eq:MSR}) as
% \begin{eqnarray}
% &&
% \lim _{ \substack{Q^2, n \rightarrow \infty \\ Q^2/n = M^2 } }
% \frac{ \pi (Q^2)^{n+1}}{n!} \left( - \frac{d \ }{dQ^2} \right) ^n
% \frac{1}{\pi} \int_0^\infty \frac{ {\rm Im} \Pi(s) }{ s + Q^2 } \, ds
% \nonumber
% \\
% && \hspace{1cm}
% = \int {\rm Im} \Pi(s) \, e^{-\frac{s}{M^2} } ds
% \nonumber
% \ \ ,
% \end{eqnarray}
and then, taking the same limits on the right-hand side in Eq.~(\ref{eq:MSR}),
we find {\it the exponential or Borel sum rule}:
\begin{eqnarray}
\M_{\text{OPE}}^J (M^2) = \int {\rm Im} \tilde \Pi^\scJ (s) \, e^{-\frac{s}{M^2} } ds
\label{eq:ESR}
\ \ .
\end{eqnarray}
Equation (\ref{eq:M0}) expresses the Borel transform of the correlator,
by which a term scaling as $( 4m^2 + Q^2 )^{-d} \cdot \langle {\mathcal O} ^{(d)} \rangle$ in the OPE (\ref{eq:OPE0})
is transformed to be %$ M^{ - (d-2) } e^{-4m^2/M^2} / (d/2-1)! \, \langle {\mathcal O} ^{(d)} \rangle$.
 $ \{ \,  M^{  (d-2) } (d/2-1)!  \, \}^{-1}  \langle {\mathcal O} ^{(d)} \rangle \, e^{-4m^2/M^2} $.
Therefore, the exponential sum rule (\ref{eq:ESR}) scales by the {\it Borel mass} $M^2$ as
\begin{eqnarray}
\sum_d \, \frac{  M^{ - (d-2) }  }{ \, ( d/2 -1) ! \,  } \, \langle {\mathcal O^{(d)} } \rangle
\sim  \int {\rm Im} \tilde \Pi^\scJ (s) \, e^{-\frac{s - 4m^2 }{M^2} } ds
\label{eq:ESR1}
.
\end{eqnarray}
Now, it is evident that the lowest pole contribution is efficiently separated owing to an
exponential factor suppressing the excited states and continua for a
small Borel mass $M^2$,
whereas the series representation by the OPE is better convergent
when $M^2$ is sufficiently large. The charmonium mass spectrum is
reliably obtained from the QCDSR since
%as we will see in Sec.~\ref{sec:QCDSR},
one can find an intermediate band of the Borel
mass $M^2$ called the ``{\it Borel window}'' in which the above
requirements, convergence and separation, are compatible to each other \cite{Bertl}. 
Analysis of the Borel window in the exponential sum rule
(\ref{eq:ESR}) is simpler and can be done in a more systematic way than
those with the moment sum rule containing two parameters $Q^2$ and
$n$. Note also that the Wilson coefficients are transformed to be
suppressed by a factorial of the operator dimension $1/(d/2-1) ! $, and 
thus the convergence of the OPE is improved. Therefore, we will use the exponential sum rule
(\ref{eq:ESR}) in subsequent sections.

%%%%%%%%%%%%%%%%%%%%%%%%%%%%%%%%%%%%%%%%%%%%%%%%%%%%%%%%%%%%%%%
%%%%%%%%%%%%%%%%%%%%%%%%%%%%%%%%%%%%%%%%%%%%%%%%%%%%%%%%%%%%%%%

\section{Spectral ansatz in the presence of mixing effects \cite{Letter}}

\label{sec:phen}

\begin{figure*}
 \centering
 \includegraphics[width=2\columnwidth]{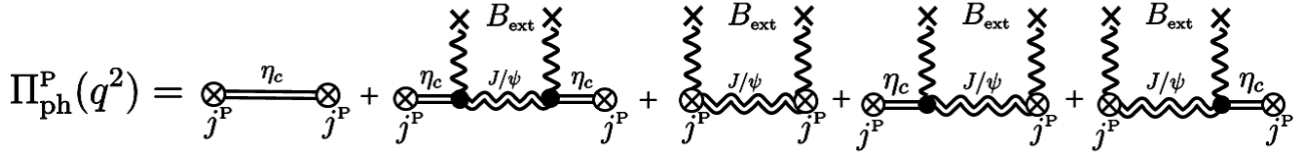}
 \caption{Diagramatic representation of the phenomenological side for the pseudoscalar channel.}
 \label{fig:phen}
\end{figure*}

\begin{figure}[t]
 \centering
 \includegraphics[width=\columnwidth]{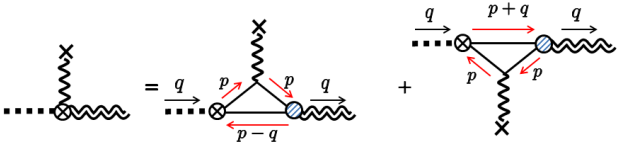}
 \caption{A direct-coupling strength between the pseudoscalar (vector) current 
 and $\eta_c$ ($\Jp$) from triangle diagrams. 
Shaded vertices show form factors given by the Bethe-Salpeter amplitudes 
of the S-wave quarkonia (\ref{eq:G5}) and (\ref{eq:Gvec}), 
while vertices with crosses denote the currents.}
 \label{fig:dir}
\end{figure}

As described in the last section, 
the current correlator \eqref{eq:corr} can be expressed in two ways;
the OPE in the deep Euclidean region ($Q^2 = -q^2 \gg1$) 
and the spectral density $\rho^J(s) = \text{Im}\tilde\Pi^J (s)/\pi$ in the physical region ($q^2>0$). 
They are connected to each other through a dispersion relation (\ref{eq:disp0}).
%that is rewritten as 
%\begin{equation}
% \tilde\Pi^J(Q^2) = \int ds \frac{\rho^J(s)}{s+Q^2} 
%  +  (\text{subtraction})
%  \label{eq:disp}
% \ .
%\end{equation}
% where we suppressed the subtraction terms which will vanish 
% due to an infinite number of derivatives operated in the Borel transformation (\ref{eq:Mn}). 
The right-hand side of Eq.~(\ref{eq:disp0}) is conventionally called ``phenomenological side'' 
because the spectral density is parametrized in hadronic degrees of freedom. 
The spectral density $\rho^J(s)$ is often assumed to have 
a perturbative continuum ${\rm Im}\, \tilde \Pi_{\rm pert}^J (s)/\pi$ 
%with a leading $\alpha_s$ correction common to the OPE and 
%
%%Comment : for light meson sector \alpha_s correction to perturbative
%%part is often neglected.
%%
and a single pole at the ground-state mass $\delta( s - m^2_{\text{pole}} )$. 
This ansatz works sufficiently well when the ground-state pole is well separated 
from a threshold of continuum as only the low-energy structure is important 
for the exponential sum rule (\ref{eq:ESR}) owing to the exponential
suppression of the higher energy part of the spectral density by the Borel transformation. 
Thus, this simple ansatz works well for the tightly bound ground-state charmonia. 
One should, however, be careful to this point in the presence of the magnetically-induced mixing 
discussed in Sec.~\ref{sec:hadron}, because it induces a $\eta_c$ (longitudinal $J/\psi$) pole 
in the longitudinal vector (pseudoscalar) current correlator. 
Therefore, there would appear two adjacent poles in the low-energy region
around the ground-state pole, and they could contribute to the
Borel-transformed correlator with the same order of magnitudes. 
We will find that an appropriate ansatz in the presence of external magnetic fields 
has a form\footnote{The ``pole+continuum'' part in the vector channel is assumed 
for a scalar part $\tilde \Pi^\V$ obtained from a spin-projection by 
the polarization vectors (\ref{eq:eps1}) and (\ref{eq:eps2}) 
and the normalization specified below Eq.~(\ref{eq:disp0}) as 
$q^{-2} \epsilon_\mu \Pi^{\V , \mu\nu} \epsilon_\nu
=  q^{-2} \epsilon_\mu \epsilon_\nu (q^\mu q^\nu - q^2 g^{\mu\nu}) \, \tilde \Pi^\V 
= \tilde \Pi^\V $ for a static charmonia.
}
\begin{eqnarray}
&&\hspace{-0.5cm}
\rho^J (s) = \pi^{-1} \left[ \, f_0 \delta( s - m^2_{c \bar c} ) 
\right.
\nonumber
\\
&&  \left.\hspace{1.7cm}
+ \theta( s - s_0 ) {\rm Im } \, \tilde \Pi^J_{\rm pert} (s) 
+ {\rm Im}\tilde \Pi^{J, \ext}_\ph (s) \,  \right ]
\label{eq:ansatz_B}
\end{eqnarray}
where $f_0$ is a coupling strength between the heavy-quark current and 
the ground-state charmonium in vacuum which is related to the mass and
electronic decay width and found to be $0.542$GeV$^2$ for the vector current
~\cite{RRY81,Lee:2013dca}. 
An effective threshold $s_0$ is fitted in the QCDSR analyses. 
The last term takes into account effects of magnetic fields as shown below. 

We carefully examine the magnetically-induced term ${\rm Im}\Pi^{J, \ext}_\ph (s) $ in the spectral ansatz. 
While we will describe calculations for the pseudoscalar channel, 
the same calculations are straightforwardly applied to the longitudinal component in the vector channel. 
A low energy expression of the pseudoscalar current correlator \eqref{eq:corr} 
in the second order of $eB$ is diagrammatically represented in Fig.~\ref{fig:phen}. 
The first diagram, surviving in the vanishing field limit, 
corresponds to the ground-state pole in Eq.~(\ref{eq:ansatz_B}), 
which is the $\eta_c$ pole in case of the pseudoscalar channel. 
As shown by the other magnetically-induced diagrams, 
we have not only an $\eta_c$ pole but also 
a longitudinal $J/\psi$ pole mixed into the pseudoscalar channel. 
One also finds that the longitudinal $J/\psi$ couples to the pseudoscalar current 
both {\it directly} and {\it indirectly}. 
% A direct and indirect coupling are induced by a three-point vertex 
% among a pseudoscalar current, an external magnetic field and a longitudinal $J/\psi$ 
% and among a $\eta_c$, an external magnetic field and a longitudinal $J/\psi$, respectively. 
A direct coupling is, as shown in Fig.~\ref{fig:dir}, induced by a three-point vertex 
among a pseudoscalar current, an external magnetic field and a longitudinal $J/\psi$. 
An indirect coupling is obtained by replacing the pseudoscalar current in the direct coupling 
by a $\eta_c$, resulting in the hadronic coupling \eqref{eq:L_pv} as already discussed. 
Thus, the second (third) diagram in Fig.~\ref{fig:phen} shows a process 
induced solely by the indirect (direct) couplings. 
Those contributions to the matrix element in Eq.~(\ref{eq:ansatz_B}) are given by 
\begin{eqnarray}
\Pi^{\ps, \ext}_\ph(q^2) &=& 
% \frac{| \langle 0|J^5| \etac \rangle|^2   }{q^2-m_{\etac}^2} +
\frac{| \langle 0|J^5| \Jp \rangle|^2   }{q^2-m_{\Jp}^2}
\ ,
\label{eq:ph}
\end{eqnarray}
% \changed{Tilde removed.}
% The matrix elements above are diagrammatically represented 
% by magnetically-induced processes in Fig.~\ref{fig:phen}. 
% While the first term in Eq.~(\ref{eq:ph}) correspond to the first diagram 
% which survives in the vanishing magnetic field limit, 
% the second term is a sum of the other diagrams in the second order of $eB$. 
with the matrix elements 
\begin{eqnarray}
| \langle 0|J^5| \Jp \rangle|^2= f_\dir 
+ f \, \frac{  | \langle P| \Jp \rangle|^2   }{(q^2-m_{\ps}^2)^2}
\ ,
\label{eq:fp}
\end{eqnarray}
where a direct-coupling strength between the pseudoscalar current and the longitudinal $J/\psi$ 
reads $f_\dir = \vert \langle V  \vert  J^5(q) \vert 0 \rangle \vert^2$, 
and a couping strength between the pseudoscalar current and a $\eta_c$ 
is proportional to $f_0$ as 
$f = \vert \langle P \vert J^5(q) \vert 0 \rangle \vert^2 = f_0 \cdot m_0^2/\pi$. 
The effective vertex \eqref{eq:L_pv} leads to 
$| \langle P| \Jp \rangle|^2 =\gamma^2$ 
%\changed{q^2/m_0^2}$, \changed{where $q^2=\omega^2$ 
for a static charmonium in the heavy-quark limit $m_0 \sim m_{\ps,\V}$. 
As shown in Fig.~\ref{fig:dir}, one can calculate the direct-coupling strength $f_\dir$ 
from two triangle diagrams 
by using the Bethe-Salpeter (BS) amplitudes of the S-wave quarkonia \cite{BSamp}. 
Led by a diagramatic calculation performed in the heavy-quark limit in Appendix~\ref{sec:mix_dir}, 
we find the direct-coupling strength as 
%$f_\dir =a_0^4 Q_c^2/64(eB)^2 f_0$
\begin{eqnarray}
f_\dir
% = \vert 2 \Phi_\dir^{(1)} \vert^2
= \frac{a_0^4 Q_c^2 }{64}  (eB)^2 f
\label{eq:dir}
,
\end{eqnarray}
with an electric charge of a charm quark $Q_c = 2/3$. 
The Bohr radius $a_0 = 0.811$ GeV$^{-1}$ is chosen to fit the
root-mean-square radius of the $\Jp$ obtained from the Cornell potential model \cite{Cornell}. 
Inserting Eq.~(\ref{eq:fp}) into Eq.~\eqref{eq:ph}, 
we find that %the second term 
the rhs in Eq.~\eqref{eq:ph} can be decomposed as 
\begin{eqnarray}
\Pi^{\ps, \ext}_\ph(q^2) &=&
\frac{ f \, \gamma^2 }{ (q^2-m_{\ps}^2)^2 (q^2-m_{\V}^2) }
\nonumber
\\
&=&
\frac{f \gamma^2}{M_-^4 }
\left[\frac{1}{q^2 - m_{\V}^2} - \frac{1}{q^2 - m_{\ps}^2}
- \frac{ M_-^2 }{ (q^2-m_{\ps}^2)^2 }  \right]
,
\nonumber
\\
\label{eq:exp}
\end{eqnarray}
\if 0
\begin{align}
\label{eq:exp}
&\hspace{-0.25cm}
% \changed{f \frac{q^2}{m_0^2}} 
\frac{ f \, \gamma^2 }{ (q^2-m_{\ps}^2)^2 (q^2-m_{\V}^2) }
% \nonumber
\\
& \ \ =
f  
\frac{\gamma^2}{M_-^4 }
\left[\frac{1}{q^2 - m_{\V}^2} - \frac{1}{q^2 - m_{\ps}^2}
- \frac{ M_-^2 }{ (q^2-m_{\ps}^2)^2 }  \right]
,
\nonumber
\end{align}
\fi
where notations are specified below Eq.~(\ref{eq:EFT}). 
Before discussing physical meaning of these terms, 
there are some comments in order. 
First, one can replace the $\Jp$ mass in the denominator by the vacuum mass $m_\V$ 
within the second-order corrections in $eB$, 
because the correlator \eqref{eq:exp} has explicit second-order corrections in the numerator. 
% \changed{Next, a dimensionless factor $q^2/m_0^2$ can be taken to be unity 
% assuming the heavy-quark limit $m_0 \sim m_{\ps,\V}$, 
% because we will take the imaginary part of the r.h.s. in Eq.~(\ref{eq:exp}), 
% resulting in delta functions. }

As for the longitudinal $J/\psi$ pole induced by the direct-coupling term, 
we find that its strength is much smaller than the hadronic-coupling strength 
of the longitudinal $J/\psi$ pole in Eq.~(\ref{eq:exp}), 
because the direct-coupling strength (\ref{eq:dir}) is proportional to the small value of 
the Bohr radius of tightly bound charmonia. 
A ratio of the direct-coupling strength over 
the hadronic-coupling strength in Eq.~(\ref{eq:exp}) 
is found to be 
\begin{eqnarray}
f_\dir / (f \gamma^2/M_-^4) \sim  0.0003 
\ ,
\label{eq:ratio}
\end{eqnarray}
so that one can safely neglect the contributions of the direct couplings. 
We also neglect cross terms depicted by the last two diagrams in Fig.~\ref{fig:phen}. 
Possible corrections to the direct-coupling strength in higher-order of $eB$ 
should be neglected in the present framework 
so that the correlator is consistently constructed within the second order in $eB$. 
Another possible correction might be a distortion of the S-wave wave function 
in an external strong magnetic field, while in the above calculation 
we inserted the Coulombic wave function in the ordinary vacuum. 
However, because of the small ratio (\ref{eq:ratio}), 
the modification of the wave function in strong magnetic fields 
would not be important, unless the wave function is very strongly distorted 
by the external magnetic fields. To estimate the order of this effect, 
we should compare the magnitudes of the Coulomb force in the potential model, $\kappa/r^2$, 
with that of an external magnetic field. 
With a strength of the Coulomb force $\kappa = 0.52$ from Ref.~\cite{Cornell} and 
the Bohr radius $a_0 = 0.811 \ {\rm GeV} \sim 0.16 \ {\rm fm}$ above, 
we have $eB / ( \kappa / a_0^2 ) \sim 0.25$ 
even for the maximal strength $eB = 10 m_\pi^2$ \cite{Bestimates}. 
This estimate indicates that distortion of the Coulombic wave function 
by the external magnetic fields will be so small that we can still neglect 
the direct coupling under the modifications of the wave function in strong magnetic fields. 
Similarly, finite temperature/density effects in the heavy-ion collisions 
could act on the direct coupling. However, these effects would be also so small, 
basically because the strength $f_\dir$ is proportional to the Bohr radius $a_0$, 
while the mixing strength between $\eta_c$ and the longitudinal $J/\psi$ 
is independent of the Bohr radius. 
Even if the Bohr radius becomes ten times larger for charmonia melting in the hot medium, 
the ratio shown in Eq.~(\ref{eq:ratio}) is still of order $10^{-3}$, 
so that one can neglect the direct coupling compared to 
the hadronic mixing between $\eta_c$ and the longitudinal $J/\psi$.

To understand physical meaning of the terms in Eq.~(\ref{eq:exp}), 
it is instructive to compare them with the second-order perturbation theory 
performed in Sec.~\ref{sec:hadron}. 
Note that, neglecting the direct couplings, 
the pseudoscalar current is first coupled to an $\eta_c$ in any process 
whether the intermediate state is an $\eta_c$ or $J/\psi$. 
Therefore, by using the coupling strength $f$, 
the current correlator may be written as 
\begin{eqnarray}
\Pi_{\rm 2nd}^\ps (q^2) =
f  \left[ \
\frac{\vert ( P \vert \etac )_{\scriptscriptstyle B} \vert^2}{q^2-m_\etac^2}
+ \frac{\vert ( P \vert \Jp )_{\scriptscriptstyle B} \vert^2}{q^2-m_\Jp^2}
\ \right]
\label{eq:2nd}
,
\end{eqnarray}
where physical masses $m_{\eta_c, J/\psi}$ %of a $\eta_c$ and longitudinal $J/\psi$ 
and corresponding wave functions in the presence of the mixing effect have been obtained 
in Eqs.~(\ref{eq:Jpsi_2nd}) and (\ref{eq:wave}), respectively. 
Now we will find that all three terms in  Eq.~\eqref{eq:exp} 
follow from an expansion of the rhs in Eq.~(\ref{eq:2nd}) 
up to the second order in $eB$. 
The first term in Eq.~\eqref{eq:exp} corresponds to putting an intermediate $J/\psi$ state on-shell 
in the second diagram in Fig.~\ref{fig:phen}. This is a production of an on-shell
$J/\psi$ from the pseudoscalar current via off-shell $\eta_c$. 
The second term with a negative sign is necessary for a conservation of 
the normalization of the spectral density, 
because the coupling of $\eta_c$ to the current must be reduced to
balance the occurrence of the coupling to $J/\psi$. 
These interpretations are confirmed by expanding the rhs in Eq.~(\ref{eq:2nd}), 
because we obtain these two terms from 
overlaps between the properly normalized unperturbed and
perturbed states, $\vert ( P \vert \etac
)_{\scriptscriptstyle B} \vert^2 \sim 1 - (\gamma/M_-^2)^2$
and $\vert ( P \vert \Jp )_{\scriptscriptstyle B}
\vert^2 \sim (\gamma /M_-^2)^2$. 
To take into account the mixing effect with maintaining the normalization, 
one should include both single poles at $\eta_c$ and $J/\psi$ 
with the residues shown in Eq.~\eqref{eq:exp}, 
giving a two-peak structure in the spectral ansatz. 
The third term has a double pole at the $\eta_c$ mass with a
factor $M_-^2$ which gives an off-shellness of a virtual $J/\psi$ in the intermediate state. 
One finds that a virtual transition to $J/\psi$ between on-shell $\eta_c$ states 
is nothing but the origin of the mass shift due to the mixing effect. 
Correspondingly, this term comes from an expansion in Eq.~(\ref{eq:2nd}) 
with respect to the mass correction of $\eta_c$ shown in Eq.~(\ref{eq:Jpsi_2nd}). 
Therefore, we have found that, if the double-pole term is included on the phenomenological side, 
it balances the corresponding %magnetically-induced mixing 
effect embedded on the OPE side 
performed on the basis of the fundamental degrees of freedom, 
and we will obtain a residual mass shift due to nonperturbative effects %other than the mixing effect 
as a result of the QCD sum rule. 
On the other hand, if the double-pole term is not included, 
we will obtain a resultant mass shift due to the mixing effect and the residual effects. 
This observation enables us to separate the residual effects of magnetic fields 
from the mixing effect, and extract effects of magnetic fields 
{\it not described in the hadronic level}. 
We will come back to this point in Secs \ref{sec:QCDSR} and \ref{sec:Bterms} 
with plots of mass shifts from QCD sum rules.

Let us perform the Borel transformation of the phenomenological side. 
Inserting the ground-state pole term in Eq.~(\ref{eq:ansatz_B}) 
into the rhs of Eq.~(\ref{eq:ESR}), we simply find 
\begin{eqnarray}
\mathcal{M}_{\text{ph}}^{\ps,\text{pole}} = f_0 e^{-m^2_{\eta_c}/M^2}
\label{eq:Mph-gs}
\ .
\end{eqnarray}
From the second term of Eq.~\eqref{eq:ansatz_B}, the Borel
transformation of the perturbative continuum part is found to be
\begin{equation}
 \mathcal{M}^{J,\text{cont}}_{\text{ph}} = \int_{s_0}^{\infty} ds
  e^{-s/M^2}\text{Im}\tilde{\Pi}^{J,\text{pert}}(s)
\end{equation}
where the expression for the perturbative continuum $\text{Im}\tilde{\Pi}^{J,\text{pert}}(s)$ is given
in Ref.~\cite{RRYrev}.
By inserting the magnetically-induced part \eqref{eq:ph} 
into the rhs of Eq.~(\ref{eq:ESR}), we obtain 
\begin{eqnarray}
\label{eq:Mph-etac}
\M_{\ph}^{\ps, \text{ext}} (M^2) &=& f_0 (eB)^2 \left[ \
Q_c^2 \frac{a_0^4}{64} e^{- \frac{m_\V^2}{M^2}}
\right.
\\
&&\hspace{-0.7cm}
\left.
+ \frac{ g_\pv^2}{M_-^4}
 \left( \, e^{- \frac{m_{\V}^2}{M^2}} - e^{- \frac{m_\ps^2}{M^2}}
+ \frac{ M_{-}^2}{ M^2} e^{- \frac{m_\ps^2}{M^2}} \, \right)
\ \right]
\nonumber
,
\end{eqnarray}
where $f_0 = \pi f / m_0^2 $ with $1/m_0^2$ coming from 
the normalization of the correlator described below Eq.~(\ref{eq:disp0}). 
% This coupling strength $f_0$ is the same as that in Eq.~(\ref{eq:ansatz_B}). 
A corresponding formula for the longitudinal $J/\psi$ can be obtained by
interchanging $m_{\ps}$ and $m_{\V}$ as
\begin{eqnarray}
\label{eq:Mph-Jpsi}
\M_{\ph}^{\V_{\scriptscriptstyle \parallel}, \text{ext}} (M^2) &=& f_0 (eB)^2 \left[ \
Q_c^2 \frac{a_0^4}{64} e^{- \frac{m_\ps^2}{M^2}}
\right.
\\
&&\hspace{-0.7cm}
\left.
+ \frac{ g_\pv^2}{M_-^4}
 \left( \, - e^{- \frac{m_\V^2}{M^2}} + e^{- \frac{m_{\ps}^2}{M^2}} 
- \frac{ \vert M_{-}^2 \vert }{ M^2} e^{- \frac{m_\V^2}{M^2}} \, \right)
\ \right]
\nonumber
.
\end{eqnarray}
The first terms in Eqs.~(\ref{eq:Mph-etac}) and (\ref{eq:Mph-Jpsi}) 
are the direct-coupling terms in Eq.~(\ref{eq:fp}) 
which are, however, negligible as discussed above. 
Following from a sign flip in $M_-^2$, we find that the double-pole contribution in the
vector channel has the opposite sign to that of the last term in Eq.~\eqref{eq:Mph-etac}.

One should remember that these magnetically-induced terms 
on the phenomenological side are not applied to the transverse $J/\psi$, 
because any magnetically-induced coupling in Fig.~\ref{fig:phen} is absent 
for the transverse component. Therefore, we will employ the conventional 
``pole+continuum'' ansatz for the transverse $J/\psi$.

These ansatz on the phenomenological side will be used to extract 
the charmonium mass spectra in Sec.~\ref{sec:QCDSR}, 
prior to which we need to examine the OPE in the next section.

%%%%%%%%%%%%%%%%%%%%%%%%%%%%%%%%%%%%%%%%%%%%%%%%%%%%%%%%%%%%%%%
%%%%%%%%%%%%%%%%%%%%%%%%%%%%%%%%%%%%%%%%%%%%%%%%%%%%%%%%%%%%%%%

\section{Operator product expansion}

\label{sec:OPE}

In this section, we include effects of a constant external magnetic field into
a series representation of the current correlator by the OPE (\ref{eq:OPE0}).
It should be noticed first that interactions between quarks 
and a constant magnetic field can be suitably regarded as a soft process,
since a constant external field does not cause any momentum transfer
to quarks in a vacuum polarization (see Fig.~\ref{fig:Pext}).
A momentum transfer is exactly zero because of a translational invariance \cite{transfer}.
A constant magnetic field is thus treated as an operator expectation value in the OPE,
and gives rise to additional terms to a series in the ordinary vacuum.
We compute the Wilson coefficients of these terms for an external magnetic field
in Sec.~\ref{sec:OPEext}, following a brief description of the OPE in the ordinary vacuum.

\subsection{OPE for charmonia in the ordinary vacuum}
%KM:  To me this subsection is redundant. 
\label{sec:OPEvac}

The OPE for heavy-quark systems was examined in detail for the
sake of investigating the lowest bound states created by the
various currents \cite{SVZ791,SVZ792, RRY80,RRY81}. As the
higher-dimension terms in the OPE are suppressed by negative
powers of the Borel mass and the factorial factors of the operator
dimension (see Eq.~(\ref{eq:ESR1})), a series representation in
the OPE is saturated by the first few terms. Indeed, it was found
that the vacuum charmonium mass spectra measured in experiments
are reproduced by including the perturbative terms and the
dimension-4 scalar gluon condensate \cite{SVZ791,RRY81,RRYrev,Bertl}
% \com{Comments on $\langle \bar c c\rangle$.}
\begin{eqnarray}
\Pi_\vac^J (Q^2) \sim C^J_{\rm p} (Q^2) \cdot {\bm 1} + C^J_{G_0} (Q^2) \cdot G_0
\label{eq:OPEvac}
\ \ ,
\end{eqnarray}
where the superscripts $J$ denote a channel of the currents (\ref{eq:jP})-(\ref{eq:jV}). 
The correlator $\Pi_\vac^\V$ and the Wilson coefficients $C^\V_{\rm p} $ and $ C^\V_{G_0}$ 
should have two Lorentz indices in the vector channel $J=V$. 
We however suppress those indices as well as 
the superscript $J$ for simplicity below as in Eq.~(\ref{eq:corr}). 
An expectation value of the scalar gluon condensate has a
form $G_0 = \langle \frac{\as}{\pi} G^{a  \mu\nu} G^a_{\mu\nu}\rangle $. 
Note that a heavy-quark condensate $\langle \bar c c\rangle$ does not contribute 
to the OPE (\ref{eq:OPEvac}) in the leading order of a heavy-quark expansion ${\mathcal O} (1/m)$, 
because it is canceled in an operator mixing with the gluon condensate \cite{SVZ792,GR84} 
(see also Sec.~3.3.5 in Ref.~\cite{RRYrev} for a comprehensive description). 
Contributions from higher-dimension gluon condensates
are small enough in the ordinary vacuum, giving stable Borel
curves \cite{RRY84}.  Such stability is maintained even at finite
temperature up to around  1.1 times the QCD phase transition temperature \cite{ML}.  
The properties of charmonium extracted
from such calculations have been recently shown to be consistent with
that obtained by solving the Schr\"odinger equation with the free
energy potential extracted from lattice calculations
\cite{Lee:2013dca}.   
However, above this temperature, 
the contributions from higher dimensional operators cannot be
neglected \cite{KL01}, and a different resummation technique will be more
appropriate to calculate the OPE \cite{Furnstahl-Hatsuda-Lee90}. 
In our present analysis, we include the terms up to the
dimension-4 scalar gluon condensate as in Eq.~(\ref{eq:OPEvac})
since recent studies have shown that effects of external magnetic
fields on the gluon condensate is sufficiently small as briefly discussed below \cite{BMW_GG,Ozaki}. 
% and the lower boundary of the Borel window is taken \com{as described in Sec.~\ref{sec:Borel} }
% so that contributions of the higher--dimension terms
% are sufficiently suppressed compared to those from the perturbative terms. 
A summary of the Wilson coefficients is available in Refs.~\cite{RRYrev,NR83}.

Following the definition of the moments (\ref{eq:Mn}),
one can straightforwardly calculate the moments of the Wilson coefficients in Eq.~(\ref{eq:OPEvac}).
This has been carried out systematically in various channels in the RRY papers \cite{RRY80,RRY81},
and explicit forms in their conventions are given by
\begin{eqnarray}
M_n^\vac = A_n ( 1 + \alpha_s \, a_n + \phi_b \, b_n)
\label{eq:Mn_vac}
\ \ .
\end{eqnarray}
An overall factor $A_n$ corresponds to the leading-order perturbative term 
in the zeroth order of the QCD coupling constant $g_s$.
The second and third terms between the parentheses give the next-to-leading order perturbative correction
and the leading power correction by the scalar gluon condensate,
which are respectively proportional to the fine structure constant in QCD, $\alpha_s = g_s^2/(4 \pi)$,
and the scalar gluon condensate,
\begin{eqnarray}
\phi_b = \frac{ 4\pi^2 G_0 }{ 9 (4m^2)^2 }
\label{eq:phib_G}
\ \ .
\end{eqnarray}
These coefficients $A_n$, $a_n$, and $b_n$ are shown in Table 1 in Ref.~\cite{RRY81}.

A useful recipe for taking the simultaneous limits $Q^2 , n\rightarrow \infty$ in Eq.~(\ref{eq:M0})
was provided in the appendix of Ref.~\cite{Bertl} with the help of a relation between special functions
(see also Appendix \ref{sec:W} in this paper).
Following the description therein, one obtains the Borel-transformed Wilson coefficients to be
\begin{eqnarray}
\M^\vac (\nu) = \pi e^{-\nu}  A(\nu) \left [ \, 1 + \as a(\nu) + \phi_b b(\nu) \, \right ]
\label{eq:M_vac}
\ \ ,
\end{eqnarray}
where a dimensionless inverse Borel mass is defined by $\nu = 4m^2/M^2$.
The coefficients in Eq.~(\ref{eq:M_vac}) correspond to those denoted by the same alphabets in Eq.~(\ref{eq:Mn_vac}).
Explicit forms of $A(\nu)$, $a(\nu)$ and $b(\nu)$ are summarized in appendices of Refs.~\cite{Bertl,ML10}.
%\com{Please put comments on differences between expressions given in those references, if necessary.}

\subsection{OPE in external magnetic fields}

\label{sec:OPEext}

\begin{figure}
 \centering
 \includegraphics[width=\columnwidth]{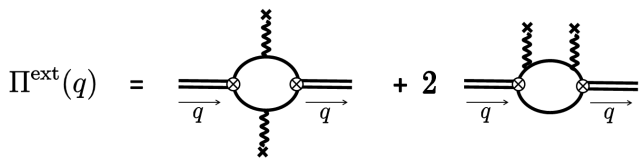}
 \caption{Diagramatic representation of corrections by an external magnetic field.
 External fields and a 1-loop perturbative part correspond to
 the operators $\langle F_{\alpha\beta} F_{\gamma\delta} \rangle$
 and the Wilson coefficients $C^{\alpha\beta\gamma\delta}$ in Eq.~(\ref{eq:Pext0}), respectively.
 }
\label{fig:Pext}
\end{figure}

We include effects of a constant external magnetic field into the
OPE as an operator for the soft dynamics. Since the magnitude of
external fields up to the LHC energy satisfies the condition
$\vert e \bm B \vert \ll 4m^2 + Q^2 $ discussed in
Sec.~\ref{sec:description}, the OPE in an external magnetic field
is thus implemented as a sum of the conventional terms in the
ordinary vacuum (\ref{eq:OPEvac}) and those from the external
magnetic field shown in Fig.~\ref{fig:Pext} as
\begin{eqnarray}
\Pi (Q^2) = \Pi^\vac (Q^2) + \Pi^\ext (Q^2)
\label{eq:Pve0}
\ \ .
\end{eqnarray}
We first remark on a possible modification of the gluon condensate 
$\langle G^{a \mu\nu} G^a_{\mu\nu} \rangle $ in the vacuum part $\Pi^\vac (Q^2) $ 
caused by external magnetic fields. 
While the light-quark
condensates in magnetic fields at zero temperature and density
have been known to increase by a mechanism called ``magnetic catalysis'' 
\cite{GMS, BMW_qbarq}, a similar growth of a gluon
condensate at zero temperature and density was recently observed
in both lattice QCD and analytic studies \cite{BMW_GG,Ozaki}.
Modification of a gluon condensate should be small, since external
magnetic fields do not directly couple to gluons, but indirectly
through sea quarks. 
Indeed, this modification is estimated to be less than 10 \% for
a magnitude of external magnetic fields around and smaller than 
the pion mass squared $\vert e \bm B \vert \alt 10 m_\pi^2$.
Thus we do not take this into account in the
present work performed at zero temperature and density, 
and effects of magnetic fields can be included as 
additional terms as in Eq.~(\ref{eq:Pve0}).

Those additional terms $\Pi^\ext(Q^2) $ for an external magnetic
field are, as mentioned below Eq.~(\ref{eq:OPE0}), suppressed by
the separation scale in heavy-quark systems. Therefore, as far as
a magnitude of an external field $\vert e \bm B \vert$ is small
enough to satisfy a hierarchy $\vert e \bm B \vert \ll ( 4m^2 +
Q^2 ) $, we can truncate a series up to dimension-4 operators
composed of a product of two field strength tensors
$F^{\mu\nu}_\ext F^{\alpha\beta}_\ext$. Beyond this separation
scale $\vert e \bm B \vert \agt ( 4m^2 + Q^2 ) $, one has to resum all
the terms being proportional to products of
arbitrary number of the field strength tensors. This resummation
can be performed by utilizing the proper-time method \cite{Sch}
which has been applied to a vector current correlator 
% as one of the authors performed recently 
(see Ref.~\cite{HI} for a recent calculation and references therein). 
Here, we examine effects of magnetic fields in a region 
$\vert e \bm B \vert \alt 10 m_\pi^2$
where  higher-order terms in $\vert eB \vert^n$ are suppressed by 
$\vert eB \vert^n / (4m^2 + Q^2)^n \sim  (10 m_\pi^2)^{n} / (4m^2 + Q^2)^n \ll 1$.
Therefore, it is sufficient to include only dimension-4 operators
without those higher-dimension operators.

Corrections by the dimension-4 operators %to leading order in $\alpha_s$ 
are diagrammatically shown in Fig.~\ref{fig:Pext} 
and have two insertions of external field lines denoted by the wavy lines.
Inserted external fields and a one-loop part correspond to
an expectation value of the dimension-4 operator and the Wilson coefficient, respectively.
We have three diagrams in total.
One is a diagram with an insertion on the each quark line (first diagram),
and the other two are diagrams with two insertions on either of the quark lines (second diagram).
The latter two diagrams provide the same contributions, resulting in a factor of 2.

One of suitable gauges for computation of the diagrams in external fields is the Fock-Schwinger gauge
also known as the fixed-point gauge \cite{Sch,NSVZrev,NR83}.
Within this gauge, a gauge field for an external constant field is expressed by
the field strength tensor $A^\mu_\ext = x_\nu F^{\nu\mu}_\ext /2$,
and thus a general form of $\Pi^\ext$ in the OPE (\ref{eq:Pve0}) is decomposed into
the operator part and its coefficient, 
\begin{eqnarray}
\Pi^\ext (q) &=& C^{\alpha\beta\gamma\delta} \cdot \frac{\alpha_\EM}{\pi} \langle F_{\alpha\beta} F_{\gamma \delta} \rangle
\label{eq:Pext0}
\end{eqnarray}
where the Wilson coefficient $C^{\alpha\beta\gamma\delta}$ has Lorentz indices
resulting from a trace of the gamma matrices,
and the fine structure constant is defined by $\alpha_\EM = e^2/(4\pi) \sim 1/137$.
% Note that $\Pi_\ext $ and thus $\pi^{\alpha\beta\gamma\delta}$,
% $\pi_0$ and $\pi_2^{\alpha\beta}$ are supposed to have additional two Lorentz indices
% in cases of the vector and the axial--vector channels.
Note again that the correlator $\Pi^\ext $ and thus $C^{\alpha\beta\gamma\delta}$
are supposed to have additional two Lorentz indices
in case of the vector channel,
which are suppressed for simplicity. 
% This term contribute to OPE of the current correlator as a term at $d=4$ in Eq.~().

Calculation for the Wilson coefficient in Eq.~(\ref{eq:Pext0})
can be performed in the same way as that for the dimension-4 (color singlet) gluon condensate
$ \langle \frac{\alpha_s}{\pi}  G^a_{\mu\nu} G^{a \, \mu\nu} \rangle$
up to a color factor and replacement of the coupling constants.
% Since the Wilson coefficients of the dimension-4 gluon condensates 
%in the pseudoscalar and vector channels 
% have been calculated for the scalar condensate \cite{RRY80, RRY81} 
% and the twist-2 condensate \cite{KKLW, SLM}, 
Since the dimension-4 scalar gluon condensate \cite{RRY80, RRY81} 
and twist-2 gluon condensate \cite{KKLW, SLM} have been known for some time ,
we confirm and apply them after taking care of the color factors. 
To make use of the preceding calculations,
it is useful to decompose the tensor structure in Eq.~(\ref{eq:Pext0}).
%Because the tensor structure of $C^{\alpha\beta\gamma\delta}$ should be constructed
%by tensors in hand, that is, an external momentum $q^\mu$ and the metric tensor $g^{\mu\nu}$,
An antisymmetric property of the field strength tensor leads to
decomposition of the right-hand side in Eq.~(\ref{eq:Pext0}) as
\begin{eqnarray}
\Pi^\ext (q) &=&
\frac{\alpha_\EM}{\pi}  [ \, C_0 \, \langle F_{\alpha\beta} F^{\alpha\beta} \rangle
+  C_2^{\alpha\beta} \, \langle F_{\alpha\gamma} F_{\beta}^{\ \gamma} \rangle_{\rm TS}  \, ]
\label{eq:Pext1}
\ \ ,
\end{eqnarray}
where a subscript ``TS'' denotes the traceless symmetic part.
Parity-odd operators $\langle F_{\alpha\beta} \tilde F^{\alpha\beta} \rangle $
and $\langle F_{\alpha\gamma} \tilde F_{\beta}^{\ \gamma} \rangle  $ would contribute
if $C^{\alpha\beta\gamma\delta}$ contained the completely antisymmetric tensor $\epsilon^{\mu\nu\sigma\rho}$.
However, this is not the case
because $C^{\alpha\beta\gamma\delta}$ is a parity-even quantity containing an even number of $\gamma^5$.
The first and second terms would be called scalar and twist-2 terms as in the case of gluon condensates, respectively.
While the twist-2 gluon condensate vanishes in the ordinary vacuum because of Lorentz symmetry,
we have a non-vanishing contribution of the twist-2 term for external fields in Eq.~(\ref{eq:Pext1})
because externally applied electromagnetic fields break Lorentz symmetry
as in the cases of finite temperature and/or density 
% \cite{KKLW, KL01, SLM, LMS, ML, ML08, ML10, ML12, GMO}.
\cite{KKLW, KL01, SLM, ML, ML08, ML10}.
With the decomposed form (\ref{eq:Pext1}), we can apply preceding calculations 
of the Wilson coefficients as shown below.

Prior to going into explicit forms of the Wilson coefficients,
let us specify a configuration of an external magnetic field.
An expectation value of the field strength tensor is given by that of an externally applied classical field
\begin{eqnarray}
\langle F^{\alpha \beta } \rangle = F^{\alpha\beta}_\ext
\ \ .
\end{eqnarray}
Here, we assume an external magnetic field extending into the positive third direction, 
of which field strength tensor is specified by $F^{21}_\ext = - F^{12}_\ext = B$ %$F^{30} = - F^{03} = E$
with all the other vanishing elements. 
In this configuration, an expectation value of the scalar operator in Eq.~(\ref{eq:Pext1}) reads
\begin{eqnarray}
F_0 := \frac{\aem}{\pi}  \langle F_{\alpha\beta} F^{\alpha\beta} \rangle
= \frac{\aem}{\pi} \cdot 2 B^2
%= \frac{\aem}{\pi} \cdot 2(B^2-E^2)
\label{eq:F0}
\ \ .
\end{eqnarray}
As for the twist-2 operator, we find
\begin{eqnarray}
\frac{\aem}{\pi}  \langle F^{\alpha \gamma} F^{\beta}_{\ \, \gamma} \rangle_{\rm TS}
&=& \frac{\aem}{\pi} \left ( F^{\alpha \gamma}_{ \exts } F^{\ \beta}_{ \exts \, \gamma}
- \frac{1}{4}  F^{\delta\gamma}_\exts F_{\delta\gamma}^\exts  g^{\alpha\beta} \right)
\nonumber
\\
&=& F_2 ( g_\parallel^{\alpha\beta} - g_\perp^{\alpha\beta})
\label{eq:FF2}
\ \ ,
\end{eqnarray}
with an operator expectation value
\begin{eqnarray}
F_2 = \frac{\aem}{\pi} \left( - \frac{1}{2} B^2 \right)   %\left( - \frac{1}{2} \right) (B^2 + E^2)
\label{eq:F2}
\ \ .
\end{eqnarray}
A tensor structure in Eq.~(\ref{eq:FF2}) is expressed by the metric tensors
in the longitudinal and transverse subspaces
$ g_\parallel^{\mu\nu} = {\rm diag} (1,0,0,-1) $ and $ g_\perp^{\mu\nu} = {\rm diag} (0, -1, -1, 0)$,
where the directions are meant with respect to the external magnetic field.

Including the external-field part $\Pi^\ext$ in Eq.~(\ref{eq:Pext1}),
we can write down a corresponding part in the moment $M_n = M_n^\vac + M_n^\ext $ as
\begin{eqnarray}
M_n^\ext = A_n (\,  \phi_b^\ext b_n^\ext + \phi_c^\ext c_n^\ext \, )
\label{eq:Mn_ext}
\ \ .
\end{eqnarray}
The leading-order perturbative part $A_n$ is, as in
Eq.~(\ref{eq:Mn_vac}), extracted as an overall factor, and the
first and second terms correspond to the scalar and twist-2
terms. In those terms, the magnetic field strengths are included as
\begin{eqnarray}
\phi_b^\ext &=&   \kappa \frac{ 4\pi^2  F_0 }{  9 (4m^2)^2 }
= \frac{ Q_c^2 }{ 12 }  \left( \frac{eB}{m^2} \right)^2
\label{eq:pb}
\ \ ,
\\
\phi_c^\ext &=&  \kappa \frac{ 4\pi^2 F_2 }{  3 (4m^2)^2 }
= - \frac{ Q_c^2 }{ 16 } \left( \frac{eB}{m^2} \right)^2
\label{eq:pc}
\ \ .
\end{eqnarray}
% \begin{eqnarray}
% \phi_b^\ext &=&   \gamma_\EM \frac{ 4\pi^2  F_0 }{  9 (4m^2)^2 }
% = \frac{ Q_c^2 }{ 12 }
% \left[ \left( \frac{eB}{m^2} \right)^2 -  \left( \frac{ eE}{m^2} \right)^2  \right]
% \\
% \phi_c^\ext &=&  \gamma_\EM \frac{ 4\pi^2 F_2 }{  3 (4m^2)^2 }
% = - \frac{ Q_c^2 }{ 16 }
% \left[  \left( \frac{eB}{m^2} \right)^2 +  \left( \frac{ eE}{m^2} \right)^2  \right]
% \end{eqnarray}
One should note that the definitions (\ref{eq:pb}) and
(\ref{eq:pc}) are the same as those for the scalar and twist-2
gluon condensates (see Eq.~(\ref{eq:phib_G}) and also, e.g.,
Eqs.~(13) and (14) in Ref.~\cite{ML10}), up to a color factor $
\kappa = Q_c^2 \cdot \tr[{\bm 1}_{\rm color}]/\tr[t^a t^a] = 6
Q_c^2$ with the Gell-Mann matrix $t^a$ normalized to be $ \tr
[t^a  t^b] = \delta^{ab}/2$ and the electric charge of charm quark
$Q_c = 2/3$ in the unit of $\vert e \vert$.
% The rightmost expressions in Eqs.~(\ref{eq:pb}) and (\ref{eq:pc}) indicate that
% the magnitude of an external magnetic field is suppressed by a large value of charm quark mass.
% While \com{Comments on resummation}
Since this color factor has been already taken into account in Eqs.~(\ref{eq:pb}) and (\ref{eq:pc}),
we find a correspondence between the Wilson coefficients for the scalar gluon condensate $\phi_b$ in Eq.~(\ref{eq:Mn_vac})
and external field $\phi_b^\ext$ in Eq.~(\ref{eq:Mn_ext}) to be
\begin{eqnarray}
b_n^\ext = b_n
\label{eq:bb}
\ \ .
\end{eqnarray}
Thus, the moment $b_n^\ext$ for an external magnetic field
is the same as that for the scalar gluon condensate $b_n$ summarized in Table 1 in Ref.~\cite{RRY81}.
% The other Wilson coefficient $c_n^\ext$ for the twist--2 term will be examined in the next section.

% \subsubsection{The twist--2 Wilson coefficients}

% \label{sec:C2}

We shall proceed to examining the last piece $c_n^\ext$ %$c^\ext(\nu)$
from the twist-2 Wilson coefficients.
The general forms of the Wilson coefficients for the twist-2 gluon condensate
were calculated both in the pseudoscalar and the vector channels \cite{KKLW}.
% and the scalar channel \cite{SLM}.
% Since the general form in the axial--vector channel has not been provided so far,
% we have calculated and provide it in Appendix~\ref{sec:WA}.
One can apply those expressions %for the twist--2 gluon condensate
to the present cases in external magnetic fields by replacing the expectation value of the operator as
\begin{eqnarray}
 \left\langle \frac{\as}{\pi} G^{\alpha \gamma} G^{\beta}_{\ \, \gamma} \right\rangle_{\rm TS}
\rightarrow
\kappa \frac{\aem}{\pi}  \left\langle F^{\alpha \gamma} F^{\beta}_{\ \, \gamma} \right\rangle_{\rm TS}
\label{eq:rep}
\ \ ,
\end{eqnarray}
where the configuration on the right-hand side was specified in Eq.~(\ref{eq:FF2}).
After making the replacement above, one performs the Borel transform.

% The twist--2 term $\Pi_2$, that is the second term in Eq.~(\ref{eq:Pext1}), will be given below,
Below, the twist-2 term $\Pi_2 = C_2^{\alpha\beta} \,
\langle F_{\alpha\gamma} F_{\beta}^{\ \gamma} \rangle_{\rm TS}$ in Eq.~(\ref{eq:Pext1})
will be calculated and then Borel-transformed through Eqs.~(\ref{eq:Mn}) and (\ref{eq:M0}).
% resulting in the Borel--transformed Wilson coefficient $c^\ext(\nu)$ in Eq.~(\ref{eq:M}).
Those results will be represented by
a longitudinal momentum $q_\parallel^\mu = (q^0,0,0,q^3)$,
transverse momentum $q_\perp^\mu = (0,q^1,q^2,0)$,
dimensionless momentum square $\xi = y/4 = Q^2/(4m^2) $
and the Feynman integrals
\begin{eqnarray}
%J_n(y) = \int_0^1 \, \{ \, 1 + x(1-x) y \, \}^{-n} \, dx
J_n(y) = \int_0^1 \, \frac{1}{\{ \, 1 + x(1-x) y \, \}^{n}} \, dx
\label{eq:Fey}
\ \ .
\end{eqnarray}

%%%%%%%%%%%%  Pseudoscalar channel %%%%%%%%%%%%%%%%%%%%

{\it Pseudoscalar channel}.--- First, we compute the twist-2
Wilson coefficients for the pseudoscalar current (\ref{eq:jP}), of
which general form has been given for the gluon condensate in
Eq.~(9) in Ref.~\cite{KKLW}. Carrying out the replacement
(\ref{eq:rep}) in the expression therein, we obtain the twist-2
term in Eq.~(\ref{eq:Pext1}) as
\begin{eqnarray}
\Pi^P_{2} &=& \left( \frac{4\pi^2}{3} \right)^{-1}  \!\!\!\! \phi^\ext_c
\, (q_\parallel^2 - q_\perp^2 ) \, \xi^{-2} \chi^P
\label{eq:P2P}
\\
\chi^P &=& \frac{1}{2} + \frac{1}{3} \left( 1 - y \right) J_1 - \frac{1}{6} J_2 - \frac{2}{3} J_3
\ \ .
\end{eqnarray}
As mentioned below Eq.~(\ref{eq:disp0}), 
we define a dimensionless correlator $\tilde \Pi_2 = \Pi_2/q^2$ for the pseudoscalar channel,
whose expression, for a static charmonium carrying a vanishing spatial momentum $q=(\omega,0,0,0)$, 
is found to be
\begin{eqnarray}
\tilde \Pi^P_{2}  &=& q^{-2} \, \Pi^P_{2}
= \left( \frac{4\pi^2}{3} \right)^{-1}  \!\!\!\! \phi^\ext_c \, \xi^{-2} \chi^P
\label{eq:P_ps}
% \ \ ,
% \\
% \tilde \Pi^S_{2} &=& q^{-2} \, \Pi^S_{2}
% = \left( \frac{4\pi^2}{3} \right)^{-1}  \!\!\!\! \phi^\ext_c \, \xi^{-2} \chi^S
\ \ .
\end{eqnarray}
% Performing the Borel--transform of the above expressions,
% we obtain the twist--2 Wilson coefficient $c^\ext(\nu)$ as shown in Appendix~\ref{sec:W}.

%%%%%%%%%%%%% Vector %%%%%%%%%%%%%

{\it Vector channel}.---
Next, we examine the twist-2 term for the vector current (\ref{eq:jV}),
of which a general tensor form has been given in Eq.~(7) in Ref.~\cite{KKLW}.
Since the vector current correlator has the two Lorentz indices,
we will project them onto the longitudinal and transverse components
corresponding to the spin polarization states of a vector meson $J/\psi$.
While mass spectra of those spin states are degenerated
in cases of static charmonia at finite temperature and/or density,
a longitudinal polarization is distinguished
from the other two transverse polarizations in external magnetic fields.

Carrying out the replacement (\ref{eq:rep}) in Eq.~(7) of Ref.~\cite{KKLW}
and contracting the Lorentz indices between the operator (\ref{eq:FF2}) and the remaining  parts,
we obtain
\begin{eqnarray}
\Pi^{V \mu\nu}_{2} &=& \left( \frac{4\pi^2}{3} \right)^{-1}   \!\!\!\! \phi^\ext_c
\, \xi^{-2}
\\
&& %\hspace{2.5cm}
\times
\left[ - 2 \chi_1^V (P_\parallel^{\mu\nu} - P_\perp^{\mu\nu} )
+ q^{-2} ( q_\parallel^2 - q_\perp^2 ) \chi_0^V P^{\mu\nu} \right]
\nonumber
\, ,
\end{eqnarray}
where the coefficient functions are given by
\begin{eqnarray}
\chi_0^V &=&  - \frac{2}{3}  +2  J_1 - 2 J_2 + \frac{2}{3} J_3
\ \ ,
\\
\chi_1^V &=& \frac{1}{2} + \left( 1-\frac{1}{3} y \right) J_1 - \frac{3}{2} J_2
\ \ ,
\end{eqnarray}
and the projection operators are introduced as
\begin{eqnarray}
P^{\mu\nu} &=& q^2 g^{\mu\nu} - q^\mu q^\nu
\ \ ,
\\
P^{\mu\nu}_\parallel &=& q_\parallel^2 g^{\mu\nu}_\parallel - q^\mu_\parallel q^\nu_\parallel
\ \ ,
\\
P^{\mu\nu}_\perp &=& q_\perp^2 g^{\mu\nu}_\perp - q^\mu_\perp q^\nu_\perp
\ \ .
\end{eqnarray}

The spin polarizations of a vector meson are specified by 
polarization vectors:
\begin{eqnarray}
\epsilon^\mu &=& (0,0,0,1)
\label{eq:eps1}
\ \ ,
\\
\tilde \epsilon^\mu &=& ( 0, \bm n,  0 )
\label{eq:eps2}
\ \ ,
\end{eqnarray}
where $\bm n$ denotes a unit vector in the transverse plane ($\vert {\bm n} \vert = 1$).
We find simple relations $ \epsilon_\mu g_\parallel^{\mu\nu} = \epsilon^\nu$,
$\tilde \epsilon_\mu g_\perp^{\mu\nu} = \tilde \epsilon^\nu$
and $\epsilon_\mu g_\perp^{\mu\nu} = \tilde \epsilon_\mu g_\parallel^{\mu\nu} = 0$,
and some more for a static charmonium carrying $q = (\omega,0,0,0)$ as
\begin{eqnarray}
 \epsilon_\mu P^{\mu\nu} \epsilon_\nu &=&
\epsilon_\mu P_\parallel^{\mu\nu} \epsilon_\nu =
\tilde \epsilon_\mu P^{\mu\nu} \tilde \epsilon_\nu = - \omega^2
\ \ ,
\\
\epsilon_\mu P_\perp^{\mu\nu} \epsilon_\nu &=&
\tilde \epsilon_\mu P_\parallel^{\mu\nu} \tilde \epsilon_\nu =
\tilde \epsilon_\mu P_\perp^{\mu\nu} \tilde \epsilon_\nu =  0
\label{eq:11111}
\ \ .
\end{eqnarray}
Therefore, the spin projection of the dimensionless correlator is carried out
for the longitudinal polarization as
\begin{eqnarray}
\tilde \Pi^{V_\parallel }_{2 } &=&
q^{-2} \cdot \epsilon_\mu (\Pi^{V  \mu\nu}_{2 } ) \epsilon_\nu
\nonumber
\\
&=&
\left( \frac{4\pi^2}{3 } \right)^{-1} \!\!\!\!  \phi^\ext_c \,
\xi^{-2}  \left(  - \chi_0^V  + 2 \chi_1^V  \right)
\label{eq:P_long}
\ \ ,
\end{eqnarray}
and for the transverse polarization as
\begin{eqnarray}
\tilde \Pi^{V_ \perp  }_{2} &=&
q^{-2} \cdot \tilde \epsilon_\mu (\Pi^{V \mu\nu}_{2 } ) \tilde \epsilon_\nu
\nonumber
\\
&=&
\left( \frac{4\pi^2}{3 } \right)^{-1} \!\!\!\!  \phi^\ext_c \,
\xi^{-2} ( - \chi_0^V )
\label{eq:P_trans}
\ \ .
\end{eqnarray}

% We have calculated the moments $c^\ext_n$ of the correlators
% (\ref{eq:P_ps}), (\ref{eq:P_long}) and (\ref{eq:P_trans})
% according to the definition (\ref{eq:Mn}),
% and summarize the results in Appendix \ref{sec:W}.

% Now that we obtained the moments of the Wilson coefficients,
% we have the Borel-transformed coefficients as in the ordinary vacuum (\ref{eq:M_vac}). 

Now that we have the Wilson coefficients obtained in 
Eqs.~(\ref{eq:P_ps}), (\ref{eq:P_long}) and (\ref{eq:P_trans}), 
their moments and the simultaneous limits $Q^2, n \rightarrow \infty$ 
can be found straightforwardly. Similarly to the vacuum part (\ref{eq:M_vac}), 
the external-field part of the Borel-transformed correlator is found to be\footnote{
The vacuum OPE in the vector channel has been performed for 
a scalar part $\tilde \Pi_{\rm vac}^\V$ in $\Pi^{\V , \mu\nu}_{\rm vac} 
= (q^\mu q^\nu - q^2 g^{\mu\nu}) \tilde \Pi_{\rm vac}^\V $. 
This result can be applied to the present case, because we need 
the exactly same quantity $q^{-2} \epsilon_\mu \Pi^{\V , \mu\nu}_{\rm vac} \epsilon_\nu  
= \tilde \Pi_{\rm vac}^\V$ for a static $J/\psi$ in the both polarization modes specified by 
the vectors (\ref{eq:eps1}) and (\ref{eq:eps2}). This normalization is consistent with 
that on the phenomenological side (\ref{eq:ansatz_B}). 
}
% \begin{eqnarray}
% \hspace{-0.5cm}
% \M^\ext (\nu) = \pi e^{-\nu}  \! A(\nu) \left[ \,  \phi_b^\ext b^\ext (\nu) + \phi_c^\ext c^\ext (\nu) \, \right]
% \, ,
% \end{eqnarray}
% where, according to Eq.~(\ref{eq:bb}), we have
% \begin{eqnarray}
% b^\ext (\nu) = b (\nu)
% \ .
% \end{eqnarray}
% Therefore, the Borel--transformed correlator amounts to
\begin{eqnarray}
\M_{\rm OPE} (\nu) &=& \pi e^{-\nu}   A(\nu) \left[ \, 1 + \as a(\nu)
\right.
\label{eq:M}
\\
&& \left. \hspace{.2cm}
+ ( \, \phi_b + \phi_b^\ext  \,) b(\nu)  + \phi_c^\ext c^\ext (\nu) \, \right]
\nonumber
\ \ ,
\end{eqnarray}
where we have $ b^\ext (\nu) = b (\nu)$ according to Eq.~(\ref{eq:bb}),
and explicit forms of $c^\ext(\nu)$ are summarized in Appendix~\ref{sec:W}.
The Borel-transformed Wilson coefficients $A(\nu)$, $a(\nu)$ and $b(\nu)$ were obtained in Ref.~\cite{Bertl},
and are listed in appendices in Refs.~\cite{Bertl,ML10}.
% Once the remaining Wilson coefficient $c^\ext(\nu)$ is provided,
% the Borel-transformed current correlator (\ref{eq:M}) allow for investigating the charmonium mass spectra
% in various channels (\ref{eq:jP})--(\ref{eq:jA}) by the exponential rum rule (\ref{eq:ESR}).
By using the Borel-transformed OPE (\ref{eq:M}), 
we will obtain charmonium spectra in the external magnetic field in the next section.

%%%%%%%%%%%%%%%%%%%%%%%%%%%%%%%%%%%%%%%%%%%%%%%%%%%%%%%%%%%%%%%
%%%%%%%%%%%%%%%%%%%%%%%%%%%%%%%%%%%%%%%%%%%%%%%%%%%%%%%%%%%%%%%

%#! latex main.tex
\section{Results and discussions}

In this section, we show charmonium mass spectra obtained from QCD
sum rule analyses, and then examine roles of magnetically-induced
mixing terms on the phenomenological side discussed in
Sec.~\ref{sec:phen} by comparing the results with those from the
hadronic effective theory shown in Sec.~\ref{sec:hadron}. We also
investigate effects of a perturbative heavy-quark loop in an
external magnetic field as a subdominant origin of mass
modifications.

\subsection{Mass shifts from QCD sum rules}

\label{sec:QCDSR}

By means of the exponential sum rule (\ref{eq:ESR}), we will
investigate charmonium mass spectra by plugging
the phenomenological side elaborated in Sec.~\ref{sec:phen}
and all the necessary Wilson coefficients involved in the Borel-transformed correlator (\ref{eq:M}).
Accumulating the OPE (\ref{eq:M}) and the spectral ansatz on the phenomenological side
shown in Eqs.~(\ref{eq:Mph-gs})-(\ref{eq:Mph-Jpsi}),
the exponential sum rule (\ref{eq:ESR}) is expressed as ($\nu = 4m_c^2/M^2$)
\begin{equation}
 \mathcal{M}_{\text{OPE}}^J (\nu) =
  \mathcal{M}_{\text{ph}}^{J, \text{pole}}(\nu) + \mathcal{M}^{J, \text{cont}}_{\text{ph}}(\nu) +
  \mathcal{M}^{J, \text{ext}}_{\text{ph}}(\nu)
.
\label{eq:Boreldispersion}
\end{equation}
Note that the above expression is for $\eta_c$ and the longitudinal $J/\psi$
($J= P, \,  V_\parallel$)
which have the magnetically-induced terms on the phenomenological side.
Since the transverse $J/\psi$ does not have those terms,
we employ the conventional spectral ansatz %for the transverse $J/\psi$
as
\begin{equation}
 \mathcal{M}_{\text{OPE}}^{\V_{\scriptscriptstyle \! \perp} } (\nu) =
  \mathcal{M}_{\text{ph}}^{\V_{\scriptscriptstyle \! \perp} , \text{pole}}(\nu)
+ \mathcal{M}^{\V_{\scriptscriptstyle \! \perp} ,\text{cont}}_{\text{ph}}(\nu)
.
\label{eq:Boreldispersion_T}
\end{equation}
% In Eqs.~\eqref{eq:Boreldispersion} and \eqref{eq:Boreldispersion_T},
% $\mathcal{M}_{\text{ph}}^{\text{pole}}$ and $\mathcal{M}^{\text{cont}}$
% come from the conventional spectral ansatz assumed to have a single pole and continuum.
% A single-pole term $\mathcal{M}_{\text{ph}}^{\text{pole}}$ corresponds to
% the first term in Eq.~\eqref{eq:ph} and Fig.~\ref{fig:phen},
% which is given by
% \begin{eqnarray}
% \mathcal{M}_{\text{ph}}^{\text{pole}} = f_0
% e^{-m^2_{\eta_c}/M^2}
% \ ,
% \end{eqnarray}
% where $f_0$ is the (squared) coupling strength between a heavy-quark current
% and a charmonium.
% We will insert a value of this coupling strength obtained in Ref.~\cite{Lee:2013dca}.
% The other term $\mathcal{M}^{\text{cont}}$ stands for a contribution from
% a perturbative continuum %up to $\mathcal{O}(\alpha_s)$
% with a leading $\alpha_s$ correction
% $%\begin{eqnarray}
% \theta(s-s_0)\text{Im}\tilde\Pi^{\rm pert} (s)
% $ %\end{eqnarray}
% common to the OPE side \com{Is it correct?}
% where $s_0$ is an effective threshold parameter fitted in the QCDSR analyses below.
Inserting these results into the Borel-transformed dispersion relation (\ref{eq:Boreldispersion}),
the mass of the lowest-lying pole can be evaluated from an equation,
\begin{equation}
m_{c \bar c}^2(M^2) = -\frac{\partial}{\partial(1/M^2)}\ln[\mathcal{M}_{\text{OPE}}
- \mathcal{M}^{\text{cont}}_{\text{ph}}-\mathcal{M}_{\text{ph}}^{\text{ext}}]
\label{eq:m2}
\ ,
\end{equation}
where the last term on the rhs, namely the magnetically-induced term,
is understood to be absent ($\mathcal{M}_{\text{ph}}^{\text{ext}}=0$)
in the case of the transverse $J/\psi$.

\begin{figure}
 \centering
 \includegraphics[width=\columnwidth]{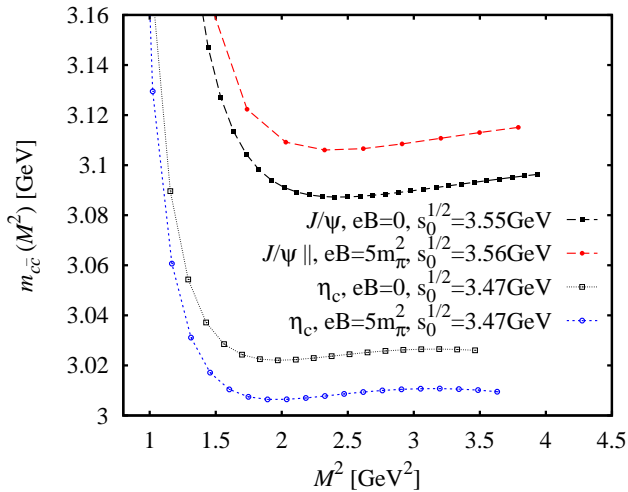}
 \caption{Borel curves for $J/\psi$ and $\eta_c$ at $eB=0$ and
 $eB=5m_\pi^2$. $s_0$ is optimized such that $m(M^2)$ is the least
 sensitive to the variation of $M^2$. }
 \label{fig:borel}
\end{figure}
Note that a mass from the QCD sum rule should be independent of
a parameter $M^2$ introduced in the Borel transformation.
Therefore, one has to examine a stability of the results with respect to
variation of $M^2$. Some examples of the $M^2$ dependence of the
charmonium masses which are obtained from Eq.~\eqref{eq:m2} and called
the Borel curves are shown in Fig.~\ref{fig:borel}.
As discussed below Eq.~(\ref{eq:ESR1}), a range of $M^2$ should satisfy two competing conditions
for a convergence of the OPE and a pole-dominance on the phenomenological side.
We require less than 30\% contribution from the dimension-4 operators to the OPE
and more than 70\% lowest-pole dominance in the dispersion integrals
(\ref{eq:Boreldispersion}) and (\ref{eq:Boreldispersion_T}), which
specifies a Borel window $M^2_{\text{min}} < M^2 < M^2_{\text{max}}$.
The effective threshold parameter $s_0$ is so tuned to make the Borel
curve the least sensitive to $M^2$. In the case of charmonia in vacuum,
the Borel curve has a minimum $m_{\text{min}}$ at
$M^2 =M^2_0 > M^2_{\text{min}}$ for $s_0=\infty$ and becomes flatter in
$M^2 > M^2_0$ as $s_0$ is decreased. Thus, we evaluate the optimized
threshold in $M^2_0 < M^2 < M^2_{\text{max}}$ for each value of the
magnetic field strength $eB$, giving the $M^2$-dependence of the mass
about and less than 10 MeV as seen in Fig.~\ref{fig:borel}.
Finally, we average the value of the mass over the same range in the Borel curve
and calculate the variance to estimate a systematic error.
Details of the systematic framework are described in Ref.~\cite{ML12}.

 \begin{figure}[!t]
  \centering
  \includegraphics[width=\columnwidth]{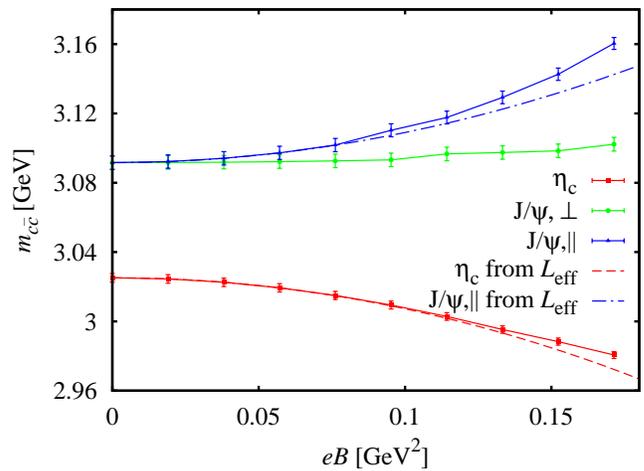}
  \caption{Mass of the charmonium states from the QCD sum rules (closed
  symbols with solid lines) and the effective Langrangian
  \eqref{eq:Jpsi_2nd} %and \eqref{eq:eta_2nd}
  (dashed line for $\eta_c$ and dash-dotted line for $J/\psi$) as functions of $eB$.}
\label{fig:mass-B}
%   \vspace{-0.2cm}
 \end{figure}

In the QCDSR analyses, we employ parameters
$\alpha_s(8m_c^2)=0.24$, $m_c(p^2=-2m_c^2)=1.26$ GeV and
$\langle \frac{\alpha_s}{\pi}G^2 \rangle = (0.35 \ \text{GeV})^4$,
and obtained the vacuum mass of $J/\psi$ and $\eta_c$
to be 3.092 GeV and 3.025 GeV, respectively.
% We use these vacuum masses as inputs
% for the masses from the effective Lagrangian \eqref{eq:Jpsi_2nd}, %and \eqref{eq:eta_2nd},
% and compare those with results from the QCDSR.
To compare results from the QCDSR with those from the effective
Lagrangian \eqref{eq:L_pv}, %\eqref{eq:Jpsi_2nd},
we insert these vacuum masses into $m_{\ps,\V}$ in Eq.~\eqref{eq:Jpsi_2nd}.
To evaluate the magnetically-induced terms on the phenomenological side,
we inserted the effective coupling $g_{\pv}=2.095$ obtained in Appendix~\ref{sec:coupling}
which was employed in Eq.~\eqref{eq:Jpsi_2nd} as well.

Figure \ref{fig:mass-B} displays the results from the QCDSR. We
first focus on $\eta_c$ and the longitudinal $J/\psi$ shown by red
and blue curves, respectively. Corresponding Borel curves at
$eB=0$ and $eB=5m_\pi^2$ are also shown in Fig.~\ref{fig:borel}.
To obtain these results, we included the phenomenological side
shown on the rhs in Eq.~\eqref{eq:Boreldispersion} for $\eta_c$
and the longitudinal $J/\psi$, but {\it not} the double-pole term
responsible for the mixing effect in
$\mathcal{M}_{\text{ph}}^{\text{ext}}$ (see
Eqs.~\eqref{eq:Mph-etac} and \eqref{eq:Mph-Jpsi}). The role of
this term and the appropriate choice of the phenomenological side
are discussed below in detail. We compare the results from the
QCDSR with those from the hadronic effective theory
\eqref{eq:Jpsi_2nd} shown by dashed and dashed-dotted lines.
Remarkably, we find a perfect agreement between the results from
the two approaches in a relatively weak-field region $eB <
0.1$GeV$^2$. This agreement indicates that the magnetically-induce
terms in Eqs.~\eqref{eq:Mph-etac} and \eqref{eq:Mph-Jpsi} are
essential ingredients to obtain physically meaningful results in
the QCDSR. In this framework, the level repulsion from the mixing
effect is simply understood as a consequence of a relative sign of
the single-pole terms. In Eq.~\eqref{eq:Mph-etac} for $\eta_c$, we
have $e^{-m^2_{\V}/M^2} - e^{-m^2_{\ps}/M^2} < 0$ owing to the
vacuum mass difference, while the corresponding terms in
\eqref{eq:Mph-Jpsi} for the longitudinal $J/\psi$ have an opposite
sign, i.e., $e^{-m^2_{\ps}/M^2} - e^{-m^2_{\V}/M^2} > 0$.
Therefore, those terms act on masses of $\eta_c$ and the
longitudinal $J/\psi$ to shift them in the opposite directions in
Eq.~(\ref{eq:m2}).

While we obtained a precise agreement in the weak-field region,
we find a slight deviation between the results from the QCDSR and the hadronic effective theory
as the magnitude of the magnetic field increases.
Moreover, we find a slight upward mass shift of the transverse $J/\psi$
shown by a green curve in Fig.~\ref{fig:mass-B},
although the transverse $J/\psi$ is not mixed with any other lowest-lying charmonium
as discussed in Sec.~\ref{sec:hadron}. Therefore, these deviations
would imply some subdominant origins of the mass shifts other than the mixing effect,
because the results from the QCDSR contain all the effects implemented in the OPE on the basis of
the fundamental degrees of freedom as well as the mixing effect in the hadronic level.
In the next section, we will argue that these effects can be separated
from the mixing effect with the help of an appropriate choice of the phenomenological side.

%%%%%%%%%%%%%%%%%%%%%%%%%%%%%%%%%%%%%%%%%%%%%%
%%%%%%%%%%%%%%%%%%%%%%%%%%%%%%%%%%%%%%%%%%%%%%

\subsection{Roles of magnetically-induced mixing terms on the phenomenological side}

\label{sec:Bterms}

\begin{figure}[!t]
  \centering
  \includegraphics[width=\columnwidth]{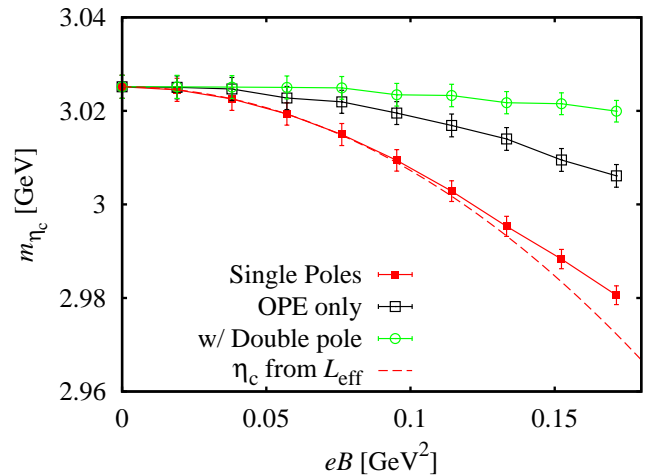}
  \caption{Mass of $\eta_c$ from the QCD sum rule with different
  implementations of the phenomenological side. }
  \label{fig:mass_phen-etac}
 \end{figure}

\begin{figure}[!t]
   \centering
   \includegraphics[width=\columnwidth]{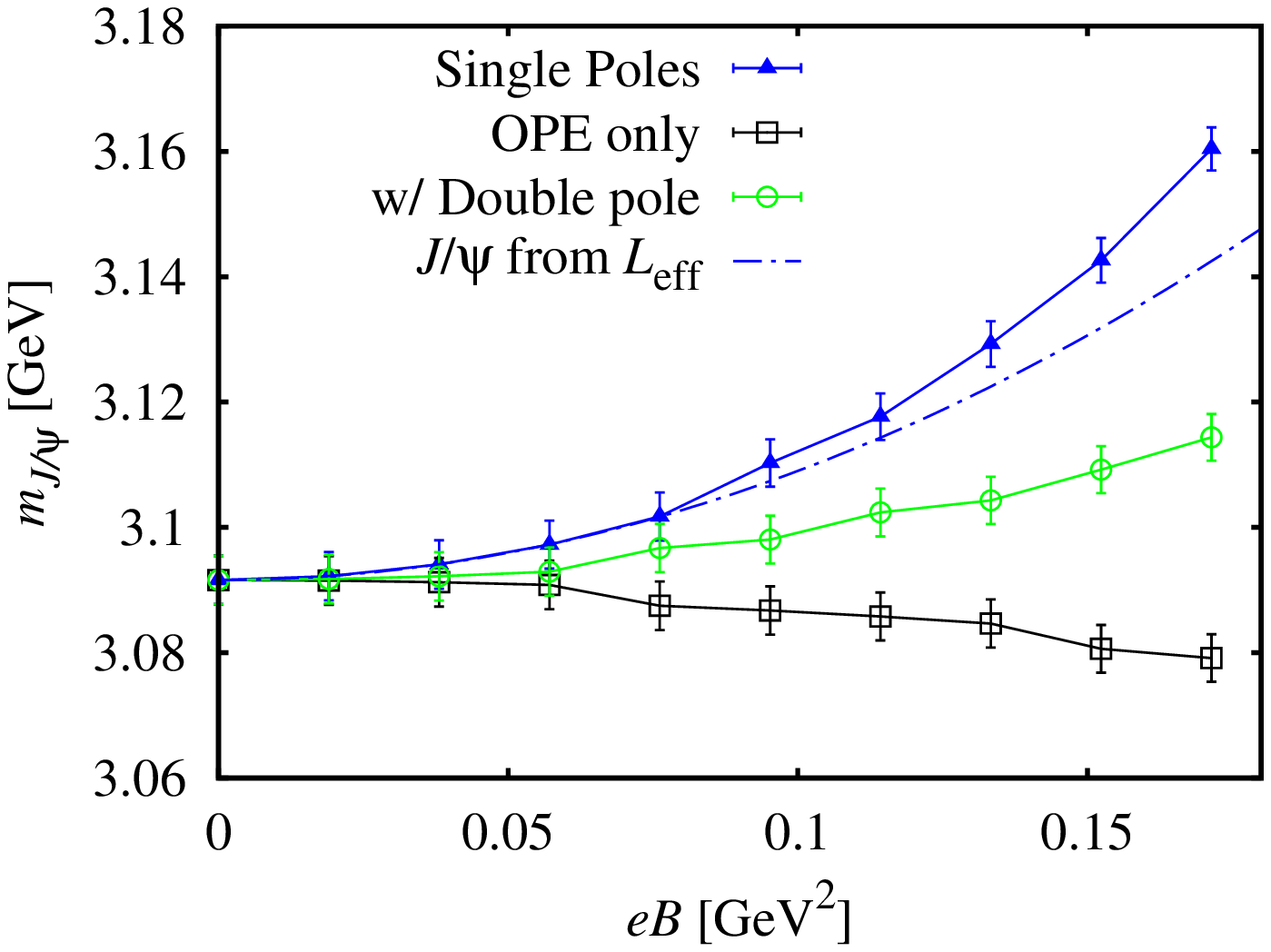}
   \caption{Mass of the longitudinal $\Jp$ from the QCD sum rule with different
   implementations of the phenomenological side. }
   \label{fig:mass_phen-jpsi}
  \end{figure}

In this section, we show QCDSR analyses in two cases
by employing (i) the conventional phenomenological side without any
magnetically-induced term $\mathcal{M}_{\text{ph}}^{\text{ext}}$
and (ii) a phenomenological side with all the terms in
$\mathcal{M}_{\text{ph}}^{\text{ext}}$ including the double-pole term
shown in Eqs.~\eqref{eq:Mph-etac} and \eqref{eq:Mph-Jpsi}.
Comparing those analyses with the one in the last section carried out with the two single poles,
we will examine a role of each term in $\mathcal{M}_{\text{ph}}^{\text{ext}}$.
In Figs. \ref{fig:mass_phen-etac} and \ref{fig:mass_phen-jpsi},
we show results for the $\eta_c$ and longitudinal $J/\psi$
in the cases (i) and (ii) with open symbols,
and the results in the last section with filled symbols.

First, the red (blue) line denoted as ``Single Poles''
in Fig.~\ref{fig:mass_phen-etac} (Fig.~\ref{fig:mass_phen-jpsi})
reminds us of the results shown in the last section where we included the single pole
of the mixing partner $J/\psi$ ($\eta_c$) in addition to the $\eta_c$ ($J/\psi$) pole,
without the double pole responsible for the mixing effect
discussed below Eq.~(\ref{eq:2nd}).
One should note that, without the double-pole on the phenomenological side,
all the information of the mass shift encoded in the OPE is reflected in the obtained masses,
while, including the double pole, the mixing effect encoded in the OPE will be
balanced and canceled by the double-pole term on the phenomenological side.
Therefore, the results without the double-pole term show
the total mass shifts including the mixing effects as well as other nonperturbative effects
from the fundamental degrees of freedom.
They are the final results from the QCDSR analysis in the present work.

Second, the black curves include neither the magnetically-induced single pole nor double pole.
In this case, obtained mass shifts would be artificial ones,
because contributions to the spectral density from
both the $\eta_c$ and longitudinal $J/\psi$ poles are attributed to
a unique pole assumed as in the conventional QCD sum rules.
This leads to an average of the $\eta_c$ and $J/\psi$ masses.
Therefore, the mass of $\eta_c$ ($J/\psi$) shown by the black curve
deviates from the red (blue) curve toward the mass of the mixing partner $J/\psi$ ($\eta_c$).
We conclude that the single pole of the mixing partner has to be included
into the spectral ansatz on the phenomenological side to subtract
the contaminating contribution from the mixing partner
and to avoid the misleading results due to the averaging.

Finally, the green curves show mass shifts obtained by including
all the terms induced by the external magnetic fields.
In this analysis, the averaging of masses discussed for the black curves
is successfully avoided by including the single pole of the mixing partner,
and the mixing effect is subtracted by including the double-pole term
which balances the corresponding contributions on the OPE side.
Therefore, the green curves show the residual mass shifts
caused by nonperturbative effects other than the mixing effect.

The roles of the magnetically-induced terms are clear now.
On the basis of the above analyses, we conclude that the dominant origin of the mass shifts
in the $\eta_c$ and longitudinal $J/\psi$ comes from the mixing between those states
as seen in comparison between the sum rule results with implementation
of the single poles (red and blue curves) and those from the hadronic
effective theory (dashed and dash-dotted lines)
and that the residual mass shifts are small in cases of charmonia.
Nevertheless, there are small mass shifts not described by the mixing effect,
and the small mass shift in the transverse $J/\psi$ shown in the last section is
not involved in the mixing effect.
We will then discuss a possible origin of these residual mass shifts
in the next section.

%%%%%%%%%%%%%%%%%%%%%%%%%%%%%%%%%%%%%%%%%%%%%%%%%%%%%%%%%%%%%%%%%%%%%%%%%%%%%

\subsection{Further mixing effects with ``continuum''}

\label{sec:SE}

\begin{figure}[t]
%\vspace{-1cm}
        \begin{center}
            \includegraphics[width=0.98\hsize]{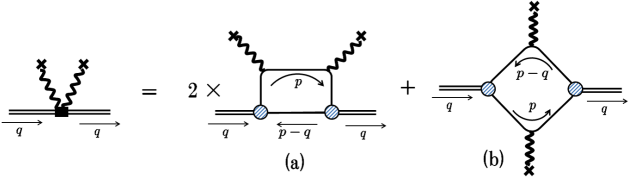}
        \end{center}
        \vspace{-0.6cm}
\caption{A heavy-quark loop as a self-energy of the charmonia in external magnetic fields.
Shaded vertices show form factors given by the Bethe-Salpeter amplitudes.}
\label{fig:SE}
\end{figure}

While we have examined mixing patterns among the charmonium states,
any other intermediate state could be contained in the physical spectral density
as long as a quantum number is matched.
Therefore, as a discussion about possible origins of the residual mass shifts found in the last section,
we shall consider interactions between an external magnetic field and
a perturbative heavy-quark loop, which are  diagrammatically shown in Fig.~\ref{fig:SE}.
As in the preceding section, we assume a static charmonium
carrying a momentum $q =  ( 2m - \eb, 0,0,0)$ with $\eb$ being the binding energy.
In Fig.~\ref{fig:SE}, a heavy quark and antiquark pair is coupled to
the heavy-quark currents with form factors given by Bethe-Salpeter amplitudes.
The Bethe-Salpeter amplitudes was obtained in
the ladder approximation and the heavy-quark limit \cite{BSamp}, %\cite{YSL,SL05},
and describes S-wave quarkonia %composed of heavy quark and anti--quark
in the ordinary vacuum.
By using the projection operators
\begin{eqnarray}
P_\pm = \frac{1}{2} ( 1 \pm \gamma^0 ) \, ,
\label{eq:prj}
\end{eqnarray}
the Bethe-Salpeter amplitudes for $\eta_c$ and $J/\psi$ are, respectively, given by
\begin{eqnarray}
\Gamma^5 (p,p-q) &=& \left(\epsilon_0 + \frac{\bp^2}{m} \right) \!
\sqrt{\frac{m_\cc}{N_c}}  \, \psi_\swave (\bp) \, P_+ \gamma^5 P_-
\label{eq:G5}
,
\\
\Gamma^\mu (p,p-q) &=& \left(\epsilon_0 + \frac{\bp^2}{m} \right) \!
\sqrt{\frac{m_\cc}{N_c}} \, \psi_\swave (\bp) \, P_+ \gamma^\mu P_-
,
\label{eq:Gvec}
\end{eqnarray}
where $\psi_\swave (\bp) $ is a ground-state wave function of the S-wave bound state
and $m_\cc$ is a mass of $\eta_c$ and $J/\psi$, which is degenerated in the heavy-quark limit.
The number of the color degrees of freedom is $N_c=3$.

We shall evaluate a self-energy of the charmonium
caused by an external magnetic field acting on a heavy-quark loop (Fig.~\ref{fig:SE}).
Since there are two diagrams (a) and (b) to be taken into account,
the self-energy is obtained as a sum of those contributions:
\begin{eqnarray}
- i \Sigma = - 2 i \Sigma^{(a)}  - i \Sigma^{(b)}
\label{eq:Sigma_sum}
\ .
\end{eqnarray}
By using the quark propagators with insertions of external magnetic fields shown in (\ref{eq:S1}) and (\ref{eq:S2}),
amplitudes of those diagrams are written down as
\begin{eqnarray}
- i \Sigma^{(a)} &=&  - \! \int \!\! \frac{d^4p}{(2\pi)^4} \tr\left[ \, \Gamma (p,p-q) S_0(p-q) \right.
\nonumber
\\
&& \left. \hspace{2.2cm} \times \, \Gamma ^\dagger (p-q,p) S_2(p) \, \right]
,
\label{eq:Sigma_a}
\\
- i \Sigma^{(b)} &=& - \! \int \!\! \frac{d^4p}{(2\pi)^4} \tr\left[ \, \Gamma (p,p-q) S_1(p-q)
\right.
\nonumber
\\
&& \left. \hspace{2.2cm} \times \, \Gamma ^\dagger (p,p-q) S_1(p) \, \right]
,
\label{eq:Sigma_b}
\end{eqnarray}
where $\Gamma$ represents the Bethe-Salpeter amplitude
(\ref{eq:G5}) or (\ref{eq:Gvec}) depending on the channels, 
and $S_0$ is the free propagator of quarks 
$S_0(p-q)=i/(\slashed{p}-\slashed{q}-m+i\varepsilon)$. 
% and $S_1$ and $S_2$ are propagators with one and two insertions of 
% external fields shown in Appendix~\ref{sec:prop_ext}, respectively. 
In the above expressions, overall minus signs on the right-hand side are
associated with a fermion loop and the QED coupling constants are
included in the propagators with external-field insertions.
% Note that the amplitudes of the loop diagrams (\ref{eq:Sigma_a}) and (\ref{eq:Sigma_b}) are gauge invariant quantities,
% while the propagators (\ref{eq:S1}) and (\ref{eq:S2}) are obtained
% in a particular gauge, that is, the Fock--Schwinger gauge.

We computed the amplitudes (\ref{eq:Sigma_a}) and (\ref{eq:Sigma_b})
in the heavy-quark limit. Following from descriptions in Appendix~\ref{sec:SE_cal},
we find
% that the amplitudes (\ref{eq:Sigma_a}) in both the pseudo--scalar and vector channels
% do not contribute to the leading order in the heavy quark limit,
% \begin{eqnarray}
% - i \Sigma^{(a)} \sim 0
% \ ,
% \end{eqnarray}
% so that the sum of the amplitudes (\ref{eq:Sigma_sum}) follows from
% the evaluation of the amplitudes (\ref{eq:Sigma_b}) as
the self energies to be
\begin{eqnarray}
\Sigma^5 \, &=& \sigma
\label{eq:sig_ps}
\ ,
\\
\Sigma^{ij} &=&  \sigma \left( \, g_\parallel^{ij}  -  g_\perp^{ij} \, \right)
\label{eq:sig_vec}
\ .
\end{eqnarray}
% A negative scalar quantity is given by
We obtained a negative scalar quantity $\sigma$ given by
\begin{eqnarray}
\sigma &=& - \frac{ ( Q_\EM B )^2  }{ 2 m^2}
\int \! \frac{d^3 \bp}{(2\pi)^3}  \, \vert \psi_\swave (\bp) \vert ^2
\left( 2 +   \frac{4m}{ \eb + \bp^2/m }  \right)
\nonumber
\ ,
\\
\label{eq:sig}
\end{eqnarray}
which contains a square of the wave function $\psi_\swave$ %of a charmonium,
and an electric charge $Q_\EM = 2/3 \vert e \vert$ and mass ``$m$''
of the quarks interacting with an external magnetic field.
We found that the self-energy in the vector channel is finite only in the spatial components $(i,j=1,2,3)$,
and that all the others vanish $(\Sigma^{00} = \Sigma^{0i} = \Sigma^{i0} =0 )$.
Since the metrics in the subspaces distinguish the longitudinal and transverse directions as introduced below (\ref{eq:F2}),
we find a mass splitting between the longitudinal and transverse modes of $J/\psi$ in external magnetic fields
as shown in a plot below.

Mass shifts of $\eta_c$ and $J/\psi$ due to the self-energies (\ref{eq:sig_ps}) and (\ref{eq:sig_vec})
can be as usual obtained from alternate insertions of the self-energies and the free propagators.
% diagrams with alternate insertions of the self--energies and the free propagators.
Inserting the self-energy $\Sigma^5$ and the free propagator %of the pseudo--scalar particle,
\begin{eqnarray}
D^5_0 (q) = \frac{i}{q^2-m_\ps^2}
\ ,
\end{eqnarray}
we obtain a resummed propagator,
\begin{eqnarray}
D^5 (q) &=& D^5_0 (q)  + D^5_0 (q)  ( -i \Sigma^5) D^5_0 (q)  + \cdots
\nonumber
\\
&=&
\frac{i}{q^2- m_\ps^2 -\sigma}
\ ,
\end{eqnarray}
and thus a mass shift of $\eta_c$ to be
\begin{eqnarray}
m_\ps^2 (B)  &=&  (m_\ps^\vac)^2 + \sigma
\label{eq:mass_ps}
\ .
\end{eqnarray}
As for $J/\psi$, inserting a free propagator in the non-relativistic limit $ ( \vert \bq \vert  \ll m_\V )$
\begin{eqnarray}
D^{ij}_0 (q) = \frac{ - i ( \, g^{ij} - q^i q^j /m_\V^2 \, ) }{q^2-m_\V^2}
\sim   \frac{ - i  g^{ij} }{q^2-m_\V^2}
\ ,
\end{eqnarray}
the resummed propagator is obtained as
\begin{eqnarray}
D^{ij} (q) &=& D^{ij}_0 (q) + D^{is}_0 (q) ( +  i \Sigma_{st }) D^{tj}_0 (q) + \cdots
% \end{eqnarray}
% Since contractions are carried out as
% \begin{eqnarray}
% + i \Sigma^s_{\ t} D^{tj}_0 (q) = \frac{ - i^2 \sigma}{ q^2 - m_\cc^2 } ( \, g_\parallel^{sj} - g_\perp^{sj} \, )
% = \frac{ + \sigma}{ q^2 - m_\cc^2 } ( \, g_\parallel^{sj} - g_\perp^{sj} \, )
% \end{eqnarray}
% we have
% \begin{eqnarray}
% \Bigl( \, + i \Sigma^s_{\ t} D^{tj}_0 (q) \, \Bigr) ^n
% = \left( \frac{ + \sigma}{ q^2 - m_\cc^2 } \right)^n \Bigl( \, g_\parallel^{sj} + (-1)^n g_\perp^{sj} \, \Bigr)
% \end{eqnarray}
% Noting a relation $g^{ij} = g_\parallel^{ij} + g_\perp^{ij} $, we obtain
% \begin{eqnarray}
% D^{ij} (q) &=&
% g_\parallel^{ij} \frac{-i}{q^2-m_\cc^2} \cdot \frac{1}{1-\frac{\sigma}{q^2-m_\cc^2}}
% + g_\perp^{ij} \frac{-i}{q^2-m_\cc^2} \cdot \frac{1}{1-\frac{- \sigma}{q^2-m_\cc^2}}
\nonumber
\\
&=&
\frac{-i  \, g_\parallel^{ij} }{q^2-m_\V^2-\sigma} + \frac{-i \,  g_\perp^{ij} }{q^2-m_\V^2+\sigma}
\ .
\end{eqnarray}
Therefore, we find the polarization-dependent mass shifts given by
\begin{eqnarray}
m_{\V \scriptscriptstyle \parallel}^2 (B)  &=&  (m_\V^{\rm vac})^2 + \sigma
\label{eq:mass_para}
\\
m_{\V \scriptscriptstyle \perp}^2 (B) &=&  (m_\V^{\rm vac})^2 - \sigma
\label{eq:mass_perp}
\ .
\end{eqnarray}
% Note that we assumed only the heavy quark limit to evaluate the self--energies,
% and therefore that the mass shifts (\ref{eq:mass_ps}), (\ref{eq:mass_para}) and (\ref{eq:mass_perp})
% can be also applied to the corresponding bottomonia.

\begin{figure}
 \centering
 \includegraphics[width=\columnwidth]{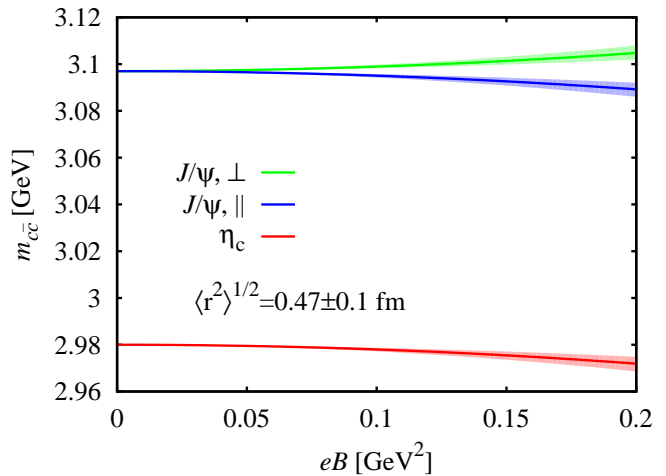}
 \caption{Mass shifts of charmonia due to self-energies.
 Masses of $\eta_c$, longitudinal $J/\psi$ and transverse $J/\psi$ are
 shown by red, blue and green curves, respectively.
 Colored stripes indicate dependences on the sizes of the bound states. }
 \label{fig:mass_r}
\end{figure}

\begin{figure}[!t]
  \centering
  \includegraphics[width=\columnwidth]{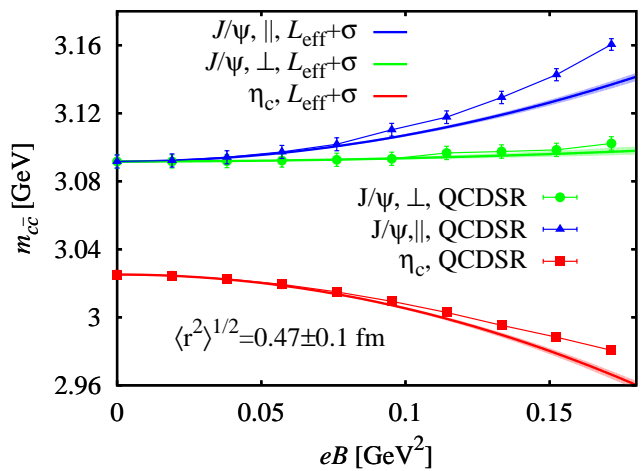}
  \caption{Masses from cooperative effects of a heavy-quark loop and a
  mixing effect, together with the results from QCDSR also shown in Fig.~\ref{fig:mass-B}.
  Colors of curves are taken to be the same as in Figs.~\ref{fig:mass-B} and \ref{fig:mass_r}.}
  \label{fig:SE_mix}
 \end{figure}

To estimate magnitudes of the mass shifts, we evaluate $\sigma$ in Eq.~(\ref{eq:sig})
assuming a Coulombic wave function,
\begin{eqnarray}
\psi_\swave (\bp) = \frac{ 8 \pi^{1/2} a_0^{ 3/2} } { ( \, (a_0 \bp)^2 + 1 \, )^2}
\label{eq:Coulomb}
\ ,
\end{eqnarray}
where the Bohr radius is related to the binding energy as $a_0^2 = ( \eb m )^{-1}$
and the wave function is normalized as
$\int \frac{ d^3 \bp }{ (2\pi)^3} \vert \psi_\s (\bp) \vert^2 = 1$.
Inserting the Coulombic wave function into Eq.~(\ref{eq:sig}), the momentum integral is carried out as
\begin{eqnarray}
\int \frac{ d^3 \bp }{ (2\pi)^3} \left( \frac{4m}{ \eb + \bp^2/m } \right) \vert \psi_\swave (\bp) \vert^2
&=& \frac{5}{2} (m a_0)^2
\ ,
\end{eqnarray}
and we obtain the $\sigma$ as a function of the Bohr radius,
\begin{eqnarray}
\sigma  &=&
- \frac{ ( Q_\EM B )^2  }{ 2 m^2}  \left( 2 +   \frac{5}{2} (m a_0)^2 \right)
\ .
\end{eqnarray}
The Bohr radius is related to a mean-square-root radius of
a Coulombic bound state as $\langle r^2 \rangle = 3 a_0^2$,
% , which follows from computing an expectation value of $r^2$ with the wave function (\ref{eq:Coulomb}).
% and inversely
% \begin{eqnarray}
% a_0 = \sqrt{\frac{\langle r^2 \rangle}{3} }
% \end{eqnarray}
where the mean-square-root radius was estimated as a typical size of the S-wave charmonium
by fitting the experimental data in terms of the Cornell potential model \cite{Cornell}.
Inserting a value $\sqrt{\langle r^2 \rangle} = 0.47 \ {\rm fm} $ obtained in Ref.~\cite{Cornell}
and the vacuum masses of charmonia into
Eqs.~(\ref{eq:mass_ps}), (\ref{eq:mass_para}) and (\ref{eq:mass_perp}),
we show the mass shifts due to the self energies in Fig.~\ref{fig:mass_r}.
Clearly, we find a mass splitting of the longitudinal and transverse $J/\psi$.
The heavy-quark loop acts to decrease the longitudinal $J/\psi$ mass,
while we have found an increasing longitudinal $J/\psi$ mass in the mixing effect.
Varying a value of $\langle r^2 \rangle$ as indicated by colored stripes,
we confirm that the magnitudes of mass shifts only weakly depend on a value of $\langle r^2 \rangle$.
To show  cooperative effects of the heavy-quark loop
and the mixing between $\eta_c$ and the longitudinal $J/\psi$,
we replace the vacuum masses in Eqs.~(\ref{eq:mass_ps}) and (\ref{eq:mass_para})
by those from the mixing effects (\ref{eq:Jpsi_2nd}).
The resultant masses are as precise as the second order in $eB$.
In Fig.~\ref{fig:SE_mix}, we find that the mixing effect overwhelms
the effect of the heavy-quark loop on the longitudinal $J/\psi$,
showing an increasing behavior of the longitudinal $J/\psi$ mass with an increasing $eB$.
Comparing Fig.~\ref{fig:mass_r} with the results from QCDSR,
we find a qualitative agreement in all three of the charmonium states.
The slightly increasing mass of the transverse $J/\psi$ is well
reproduced by the heavy-quark loop effect.
While we need more detailed information of the physical spectral density
to include resonance structures and so on, this agreement implies
that the perturbative heavy-quark loop is one of the subdominant
origins of the mass shifts in external magnetic fields.

%%%%%%%%%%%%%%%%%%%%%%%%%%%%%%%%%%%%%%%%%%%

%%%%%%%%%%%%%%%%%%%%%%%%%%%%%%%%%%%%%%%%%%%%%%%%%%%%%%%%%%%%%%%
%%%%%%%%%%%%%%%%%%%%%%%%%%%%%%%%%%%%%%%%%%%%%%%%%%%%%%%%%%%%%%%

\section{Summary}

\label{sec:summary}

We investigated effects of strong magnetic fields on the mass
spectra of $S$-wave charmonium states, i.e., $\etac$ and $\Jp$,
and elaborated the ansatz for the spectral density in the QCD sum rule
method, the so-called phenomenological side,  to consistently manipulate
mixing effects in external magnetic
fields. We implemented quadratic terms in the order of magnetic
fields for the spectral ansatz and discussed a role of each term on the
basis of a partial fraction decomposition (\ref{eq:exp}) and numerical
analyses. With an
appropriate form of the spectral ansatz obtained in the present
work, we found that the mass shifts of static $\eta_c$ and the
longitudinal $J/\psi$ precisely agree with those obtained from an
effective Lagrangian approach, indicating that the dominant effect
of magnetic fields comes from a level repulsion between those
two states. As for the transverse $\Jp$, we obtained an increasing
mass with respect to an increasing magnitude of a magnetic field,
while the transverse $\Jp$ is not mixed with any state.

This behavior of the transverse $\Jp$ and
residual mass shifts of $\eta_c$ and the longitudinal $J/\psi$
imply existence of some other effects not fully described
by the mixing effect in leading-order effective Lagrangian in mesonic degrees of freedom.
We examined effects of a mixing effect with higher states and continuum.
This was carried out by approximating the intermediate states
as a perturbative heavy-quark loop with two insertions of external magnetic fields.
We found that this effect gives rise to a splitting between the longitudinal and transverse $\Jp$
and indeed an increasing mass of the transverse $\Jp$,
while we need more precise information of the spectral density
for the higher state and continua to reach a fully conclusive result.

While the residual mass shift, other than the mixing effect, is
found to be small for the charmonia, our analysis indicates that
one has to take into account effects of the magnetic fields on the
phenomenological side consistently to the OPE side. An interesting
application would be the QCDSR analysis on light and heavy-light mesons. 
For instance, a peculiar behavior of $\rho$ 
meson spectrum observed in strong magnetic fields by lattice QCD
simulations \cite{HY, rho_lat} might be related to changes of QCD vacuum
properties in the strong magnetic field limit as mentioned in Introduction. 
As the OPE for such a light meson manifestly includes vacuum expectation values, 
e.g., a quark condensate $\langle \bar q q \rangle$, one could investigate
how the vacuum properties are reflected in light-meson spectra by the
QCDSR method, %developed with an appropriate treatment of the mixing effect, 
where one would expect a larger nonperturbative effect 
than in charmonia. The elaborate treatment of the mixing effect is necessary even in
any other methods involving the spectral density by means of the
correlation functions in constant magnetic fields. A general
framework discussed in the present work allows for extracting
nonperturbative effects of magnetic fields on QCD bound states,
and will shed light on deeper understanding of the interplay between
QCD and QED on the basis of the fundamental degrees of freedom. 

% It would be also interesting to investigate charmonium spectra in strong electric fields,
% because the early-time dynamics in the ultrarelativistic heavy ion collisions
% is accompanied by not only a magnetic field but also an electric field \cite{Bestimates}.
% In our study of mixing patterns, it turned out that mixing patters
% in electric fields are different from those in magnetic fields.

%%%%%%%%%%%%%%%%%%%%%%%%%%%%%%%%%%%%%%%%%%%%%%%%%%%%%%%%%%%%%%%
%%%%%%%%%%%%%%%%%%%%%%%%%%%%%%%%%%%%%%%%%%%%%%%%%%%%%%%%%%%%%%%

\section*{Acknowledgements}

% \textit{Acknowledgements}.---
K.H. thanks Hung-chong Kim for fruitful conversations in the early stage of this work. 
This work was supported by the Korean Research Foundation under
Grants No.~KRF-2011-0020333 and No. KRF-2011-0030621. K.M. is
supported by HIC for FAIR, the Polish Science
Foundation (NCN) under Maestro Grant No. 2013/10/A/ST2/00106 and
the Grant-in-Aid for Scientific Research on Innovative Areas from MEXT
(Grant No. 24105008). 
% KM acknowledges partial support of the Polish Science Foundation
% (NCN), under Maestro grant 2013/10/A/ST2/00106.
The research of K.H. is supported by JSPS Grants-in-Aid
No.~25287066. S.C. was supported in part by the Korean Ministry of
Education through the BK21 PLUS program. Three of the authors
(K.H., K.M. and S.O.) thank Yukawa Institute for Theoretical
Physics, Kyoto University, where a part of this work was discussed
during the YIPQS international workshop ``{\it New Frontiers in
QCD 2013}.''

%+++++++++++++++++++++++++++++++++++++++++++++++++++++++%

\appendix

%#! latex main.tex

% \section{Power counting in OPE for heavy quark systems}

% \label{sec:counting}

% \com{Semi--rigorous proof of a counting scheme $(4m^2+Q^2)^{-d/2}$. Let me think a little bit more.}

\section{Mixing strength from experimental data sets}

\label{sec:coupling}

Here we determine the coupling constant $g_\pv $ 
which gives strength of mixing effects between pseudoscalar and vector mesons. 
We calculate radiative decay widths in a reaction $J/\psi \to \gamma \, \eta_c$ 
by employing the effective vertex (\ref{eq:L_pv}), 
and read off the coupling constant by fitting the experimental data. 
%In Appendix~\ref{sec:coupling_FS}, we calculate the coupling constant $g_\pv $ 
%by using another method in terms of Bethe-Salpeter amplitudes of the S-wave charmonia \cite{BSamp}, 
%which provides a value of the coupling constant 
%consistent with the experimental data and pNRQCD calculations \cite{pNRQCD}. 

%We shall examine a transition between 
%pseudoscalar and vector mesons, $J/\psi \to \gamma \, \eta_{c} $. 
With the interaction Lagrangian (\ref{eq:L_pv}), we obtain an invariant amplitude 
\beq
\M_\pv
&=& \langle \gamma {\rm P} | \mathcal{L}_{\scriptscriptstyle \gamma {\rm P V} } | {\rm V}  \rangle \nonumber \\
&=& - \frac{e g_\pv}{m_0} \epsilon_{\mu \nu \alpha \beta} 
k_{\gamma}^{\mu} \epsilon_{\gamma}^{\nu} p_{\V}^{\alpha} \epsilon^{\beta}_{\V}
\ \ .
\eeq
% A mass of the pseudo--scalar meson is denoted as $M_\ps$, 
% and $k^\mu_\gamma $ and $\epsilon_\gam^\mu$ ($p^\mu_\V $ and $\epsilon_\V^\mu$) represent 
% a momentum and polarization vector of the photon (vector meson), respectively. 
% Summing and averaging the polarizations of the photon and the vector meson, respectively, 
A momentum and polarization vector of the photon (vector meson) are denoted as 
$k^\mu_\gamma $ and $\epsilon_\gam^\mu$ ($p^\mu_\V $ and $\epsilon_\V^\mu$), respectively. 
Summing the polarizations of the photon and averaging those of the vector meson, 
we find
\beq
\frac{1}{3} \sum_{s_{V}} \sum_{s_{\gam}} | \M_\pv |^{2}
&=&  \frac{2}{3} \left( \frac{e g_{\pv} }{ m_0 } \right)^{2} m_{\V}^{2} \, \tilde p^{2}
\ ,
\eeq
where a magnitude of the center-of-mass momentum in the final state is given by 
$\tilde p = (m_\V^2 - m_\ps^2) / (2 m_\V)$. 
%with masses of initial and final state mesons being $M_i$ and $M_f$. 
Integrating over the phase-space volume in the two-body final state, 
the decay width is then obtained to be 
\beq
\Gamma [ {\rm V} \to \gamma {\rm P} ]
&=& \frac{\tilde p}{8 \pi m_{\V}^{2}}  \frac{1}{3} \sum_{s_{\V}}  \sum_{s_{_{\gam}}} | \M_\pv |^{2} \nonumber \\
&=& \frac{1}{12} \frac{ e^2 g_{\pv}^{2} \tilde p^{3} }{  \pi m_0^{2} }
\ .
\eeq
By fitting the measured radiative decay width, 
% with the coupling constant $g_{\pv}$ being a single parameter, 
we obtain the coupling constant $g_{\pv}$ as 
\beq
% g_{\pv} &=& \sqrt{ \frac{ 12 \pi M_{\ps}^{2} \; \Gamma_{\rm exp}  [ {\rm V} \to \gamma {\rm P} ] \; }{ \tilde p^{3} } }
g_{\pv} &=&  \sqrt{ 12 \pi e^{-2} \tilde p^{ - 3 }  m_0^{2} \; \Gamma_{\rm exp}  [ {\rm V} \to \gamma {\rm P} ] \; }
\ .
\eeq
Substituting the measured value 
$\Gamma_{\rm exp}[ J/\psi \to \gamma \, \eta_{c} ] = 1.579$ keV, %1.214 keV
we obtain the coupling strength 
\beq
% g_{\scriptscriptstyle \gamma \eta_{c} J/\psi} 
g_\pv = 2.095 
\label{eq:g_pv}
\ .
\eeq
% where $e$ denotes unit electric charge. 

%%%%%%%%%%%%%%%%%%%%%%%%%%%%%%%%%%%%%%%%%%%%%%%%

\section{Mixing strengths and self-energy from the Bethe-Salpeter amplitudes}

\label{sec:prop_ext}

By using the Bethe-Salpeter amplitudes of the S-wave quarkonia (\ref{eq:G5}) and (\ref{eq:Gvec}) 
obtained in the heavy-quark limit \cite{BSamp}, 
we can also investigate interactions between those quarkonia and external magnetic fields. 
We provide a calculation of a coupling strength in the mixing between 
$\eta_c$ and the longitudinal $J/\psi$ from triangle diagrams (Fig.~\ref{fig:triangles}),
that between the pseudoscalar (vector) current and the longitudinal
$J/\psi$ ($\eta_c$) from triangle diagrams shown in Fig.~\ref{fig:dir}, 
and the self-energies of the quarkonia (Fig.~\ref{fig:SE}). 
In Appendix~\ref{sec:coupling_FS}, we will find a coupling constant 
in the mixing between $\eta_c$ and $J/\psi$, 
of which the simple expression agrees with the one obtained in the leading-order pNRQCD calculation \cite{pNRQCD} 
and of which the value is in good agreement with the one obtained 
by fitting the experimentally measured radiative decay width in Appendix.~\ref{sec:coupling}. 
In Appendix~\ref{sec:mix_dir}, we show a mixing strength between a current and a charmonium 
used for constructing the phenomenological side of the QCD sum rule in Sec.~\ref{sec:phen}. 
In Appendix~\ref{sec:SE_cal}, we describe some details in calculation of 
the self-energy of $\eta_c$ and $J/\psi$ shown in Sec.~\ref{sec:SE}.

Interactions between quarks and external magnetic fields are taken into account 
by employing the Fock-Schwinger gauge throughout this section. 
In this gauge, quark propagators with one and two insertions of constant external fields are expressed as \cite{RRYrev} 
\begin{eqnarray}
% S_0 (p) &=& \frac{i}{\slashed p - m + i \varepsilon}
% \\
S_1 (p) &=& - \frac{i}{4} Q_\EM F_{\alpha \beta} \frac{1}{(p^2-m^2+ i \varepsilon)^2} 
\label{eq:S1}
\\
&& \hspace{.5cm} \times
\left\{ \sigma^{\alpha\beta} (\slashed p + m) + ( \slashed p + m) \sigma^{\alpha\beta} \right\}
\nonumber
\ ,
\\
S_2 (p) &=&  - \frac{1}{4} Q_\EM^2 F_{\alpha \beta} F_{\mu\nu} \frac{1}{(p^2-m^2+ i \varepsilon)^5} 
\label{eq:S2}
\\
&& \hspace{.5cm} \times
(\slashed p +m) \left\{ f^{\alpha\beta \mu\nu} + f^{\alpha\mu\beta\nu} + f^{\alpha\mu\nu\beta} \right\} (\slashed p +m)
\nonumber
\ ,
\end{eqnarray}
where $Q_\EM$ denotes an electromagnetic charge of a quark 
and the gamma matrix structures are given by 
\begin{eqnarray}
\sigma^{\alpha\beta} \ &=& \frac{i}{2} [\gamma^\alpha, \gamma^\beta]
\ ,
\\
f^{\alpha\beta \mu\nu} &=& 
\gamma^\alpha ( \slashed p + m) \gamma^\beta ( \slashed p + m) \gamma^\mu ( \slashed p + m) \gamma^\nu
\ .
\end{eqnarray}

\subsection{Mixing strength between $\eta_c$ and $J/\psi$}

\label{sec:coupling_FS}

\begin{figure}[t]
%\vspace{-1cm}
		\begin{center}
			\includegraphics[width=\hsize]{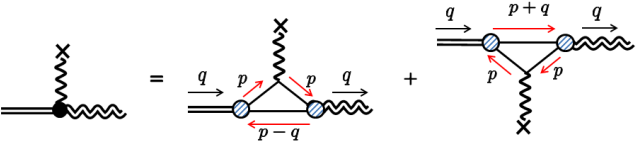}
		\end{center}
        \vspace{-0.5cm}
\caption{An effective coupling strength from triangle diagrams. 
Shaded vertices show form factors given by the Bethe-Salpeter amplitudes.
}
\label{fig:triangles}
\end{figure}

We compute the triangle diagrams in Fig.~\ref{fig:triangles} and then read off the effective coupling constant 
in the mixing between pseudoscalar and vector quarkonia in an external magnetic field. 
Calculations of the diagrams are performed in the heavy-quark limit, 
and the final result is found to be independent of the quark mass in the leading order. 
Also, note that the coupling constant is independent of the wave functions of charmonia 
although the Bethe-Salpeter amplitudes contain the wave functions 
(see Eqs.~(\ref{eq:G5}) and (\ref{eq:Gvec})).

Let us call the triangle diagrams with clockwise and counterclockwise 
ordering of vertices Diagram (a) and (b), respectively. 
We compute a sum of those diagrams 
\begin{eqnarray}
i \M^\mu  &=& i \M_a^\mu + i \M_b^\mu  
\ ,
\end{eqnarray}
which are written down as 
\begin{eqnarray}
&&
i \M_a^\mu  = - \!\! \int \! \frac{d^4p}{(2\pi)^4} 
\tr \left[ \, \Gamma_5^\dagger (p-q,p)  S_1(p) \right.
\nonumber
\\
&& \hspace{3.3cm} \left. \times
\Gamma^\mu (p,p-q)  S_0 (p-q)  \, \right]
\label{eq:M1}
\ ,
\\
&&
i \M_b^\mu  = - \!\! \int \! \frac{d^4p}{(2\pi)^4} 
\tr \left[ \,  \Gamma_5^\dagger (p+q,p) S_0 (p+q) \right.
\nonumber
\\
&& \hspace{3.3cm} \left. \times
\Gamma^\mu (p,p+q) S_1(p) \, \right]
\label{eq:M2}
\ .
\end{eqnarray}

First, we shall evaluate Diagram (a) by carrying out the momentum integral. 
One of the integrals with respect to the zeroth component can be carried out as a contour integral 
with a path enclosed either upward or downward in the complex $p^0$-plane. 
One finds that two poles are enclosed inside the contour in the each case, 
and that one of the two pole contributions is suppressed by an inverse quark mass in the heavy-quark limit. 
Enclosing the contour upward for a simplicity of the calculation, 
we pick up the leading contribution from the pole located on  
\begin{eqnarray}
\bar p^0_a = q^0 - \epsilon_{\bp-\bq} \simeq m - \left(\epsilon_0 + \frac{\bp^2}{2m} \right)
\ ,
\end{eqnarray}
where $\epsilon_0$ is a binding energy as mentioned in the beginning of Sec.~\ref{sec:SE}, 
and the approximate equality is valid in the heavy-quark limit. 
At this pole, we have a residue obtained from the limiting values, 
\begin{eqnarray}
&&
\lim_{p^0 \rightarrow \bar p_a^0} \, (p^0 - p_a^0) S_0 (p-q) 
% = \frac{ (-\epsilon_{\bp-\bq}) \gamma^0 - \bp \cdot {\bm \gamma} + m }{ - 2\epsilon_{\bp-\bq}} 
% \sim - P_-
% \nonumber
% \\
\sim - P_-
\ ,
\\
&&
\lim_{p^0 \rightarrow \bar p_a^0} S_1 (p) \sim
% - \frac{i}{4} Q_\EM F_{\alpha\beta} 
% \frac{ (2m) \left\{ \,  \sigma^{\alpha\beta} P_+ + P_+ \sigma^{\alpha\beta}  \, \right\} }
% { \left\{  -2m (\eb + \bp^2/m) \right\}^2 }
% \nonumber
% \\
% &=&
- \frac{i}{4} Q_\EM F_{\alpha\beta} 
\frac{ \sigma^{\alpha\beta} P_+ + P_+ \sigma^{\alpha\beta}   }
{ 2m (\eb + \bp^2/m) ^2 }
\ ,
\end{eqnarray}
and thus the integral in Diagram (a) is evaluated as 
\begin{eqnarray}
i \M_a^\mu  
&=& 2 Q_\EM  \tilde F^{0\mu}  \int \frac{ d^3\bp}{(2\pi)^3} \vert \psi_\swave (\bp) \vert^2
\ .
\end{eqnarray}

Similarly,% to the calculation for Diagram (a), 
we evaluate Diagram (b) by maintaining the leading pole contribution with the contour enclosed downward. 
Picking up the pole at 
\begin{eqnarray}
\bar p^0_b = - q^0 + \epsilon_{\bp-\bq} \simeq - m + \left(\epsilon_0 + \frac{\bp^2}{2m} \right)
\end{eqnarray} 
we have 
\begin{eqnarray}
&&
\lim_{p^0 \rightarrow \bar p_b^0} \, (p^0 - p_b^0) S_0 (p+q) 
% = \frac{ \epsilon_{\bp-\bq} \gamma^0 - \bp \cdot {\bm \gamma} + m }{ 2\epsilon_{\bp-\bq}} 
% \sim  P_+
% \nonumber
% \\
\sim  P_+
\\
&&
\lim_{p^0 \rightarrow \bar p_b^0} S_1 (p) \sim
% - \frac{i}{4} Q_\EM F_{\alpha\beta} 
% \frac{ (2m) \left\{ \,  \sigma^{\alpha\beta} P_- + P_- \sigma^{\alpha\beta}  \, \right\} }
% { \left\{ - 2m (\eb + \bp^2/m) \right\}^2 }
% \nonumber
% \\
% &=&
- \frac{i}{4} Q_\EM F_{\alpha\beta} 
\frac{ \sigma^{\alpha\beta} P_- + P_- \sigma^{\alpha\beta}   }
{ 2m (\eb + \bp^2/m) ^2 }
\end{eqnarray}
and then find that Diagram (b) provides the same contribution as that of Diagram (a), 
\begin{eqnarray}
i \M_b^\mu  &=& i \M_a^\mu  
\ .
\end{eqnarray}
Therefore, we obtain the sum of two triangle diagrams as 
\begin{eqnarray}
i \M^\mu & = & 2 \times 2 Q_\EM  \tilde F^{0\mu}  
\int \frac{ d^3\bp}{(2\pi)^3} \vert \psi_\swave (\bp) \vert^2
\nonumber
\\
&=& 4 Q_\EM  \tilde F^{0\mu} 
\ .
\end{eqnarray}
Note that the second line follows from a normalization of the wave function 
\begin{eqnarray}
\int \frac{ d^3\bp}{(2\pi)^3} \vert \psi_\swave (\bp) \vert^2 = 1
, 
\end{eqnarray}
and thus the amplitude is independent of the wave functions. 
We obtain the mixing amplitude between the pseudoscalar 
and the longitudinal (transverse) mode of the vector state 
by contracting with the polarization vector $\epsilon^\mu$ ($\tilde \epsilon^\mu$) 
(see (\ref{eq:eps1}) and (\ref{eq:eps2})). 
When an external magnetic field is applied in the positive third direction $(\tilde F^{30} = - \tilde F^{03} = B)$, 
we find an amplitude for the longitudinal mode as 
\begin{eqnarray}
i \M_\mu  \epsilon^\mu &=& 4 Q_\EM  \tilde F^{0\mu} \epsilon_\mu
% \nonumber
% \\
% &=& 
= 4 Q_\EM  B
\label{eq:pv}
\ ,
\end{eqnarray}
while an amplitude for the transverse modes vanishes,  
\begin{eqnarray}
i \M_\mu  \tilde \epsilon^\mu &=& 0
\ .
\end{eqnarray}

As a wrap up, we found that only the longitudinal mode of the vector state 
can mix with the pseudoscalar state in an external magnetic field, 
and that an effective vertex of the interaction among 
a photon, the pseudoscalar state and the longitudinal mode of the vector state is given by 
\begin{eqnarray}
\Lag_{\scriptscriptstyle \gamma \ps \V} = 
\frac{4 Q_\EM}{m_0} \, \tilde{F}_{\mu \nu} (\partial^{\mu}P) V^\nu
\label{eq:L_pv0}
\ .
\end{eqnarray}
The coupling constant depends only on an electric charge of a heavy quark, 
and is given by $g_{\pv} = 8/3 \simeq 2.66 $ ($g_{\pv} = 4/3 \simeq 1.33 $) 
for the transition between $\eta_c$ and $J/\psi$ ($\eta_b$ and $\Upsilon$). 
This is consistent with the value obtained by fitting the measured radiative decay width (\ref{eq:g_pv}), 
but slightly overestimated. 
We also note that we can calculate the radiative decay widths in 
$J/\psi \rightarrow \eta_c + \gamma$ and $\Upsilon \rightarrow \eta_b + \gamma$ 
by using an effective vertex (\ref{eq:L_pv0}), 
resulting in expressions consistent with those from the leading-order calculation by pNRQCD \cite{pNRQCD}. 
The overestimate mentioned above was improved 
owing to subleading terms in pNRQCD \cite{pNRQCD}.

%%%%%%%%%%%%%%%%%%%%%%%%%%%%%%%%%%%%%%%%%%%%%%%%%%%%%%%%%%%%%%%%%%%%%%%

\subsection{Direct-mixing strength}

\label{sec:mix_dir}

We compute the direct-coupling strength 
between the heavy-quark current and charmonium depicted in Fig.~\ref{fig:dir}. 
Amplitudes of theses two diagrams are written down similarly to Eqs.~(\ref{eq:M1}) and (\ref{eq:M2}) as 
\begin{eqnarray}
&&
i \M_a^\mu  = - \!\! \int \! \frac{d^4p}{(2\pi)^4} 
\tr \left[ \, i \gamma^5  S_1(p) \right.
\nonumber
\\
&& \hspace{3.3cm} \left. \times
\Gamma^\mu (p,p-q)  S_0 (p-q)  \, \right]
\ ,
\\
&&
i \M_b^\mu  = - \!\! \int \! \frac{d^4p}{(2\pi)^4} 
\tr \left[ \,  i \gamma^5  S_0 (p+q) \right.
\nonumber
\\
&& \hspace{3.3cm} \left. \times
\Gamma^\mu (p,p+q) S_1(p) \, \right]
\ .
\end{eqnarray}
We evaluate the energy integrals in the above as in the calculation in Appendix~\ref{sec:coupling_FS}, 
and obtain 
\begin{eqnarray}
\hspace{-0.5cm}
i \M_a^\mu  &=& i \M_b^\mu  
\nonumber
\\
&=&
- i Q_\EM \frac{\sqrt{N_c m_{c\bar c}}}{2m} \tilde F^{0\mu} \!\!
\int \!\!\! \frac{d^3 \bp}{(2\pi)^3}  \frac{\psi_\swave (\bp)}{ \epsilon_0 + \bp^2/m }
\ .
\end{eqnarray}
We find that only the longitudinal $\Jp$ having a polarization vector $\epsilon^\mu$ 
is directly created from the pseudoscalar current 
in the presence of external magnetic fields, 
since the above amplitude is proportional to $\tilde F^{0\mu}$. 
Assuming a Coulombic wave function (\ref{eq:Coulomb}) for $\Jp$, 
we find the direct-coupling strength as 
\begin{eqnarray}
f_\dir &=& \left| \,  - 2 i Q_\EM \frac{\sqrt{N_c m_{c\bar c}}}{2m} 
\frac{\tilde F^{0\mu} \epsilon_\mu }{4 \pi^{1/2} a_0^{3/2} \epsilon_0^{1/2} }
\, \right|^2
\nonumber
\\
&=&
f \frac{a_0^4 ( Q_\EM B)^2}{64}
\end{eqnarray}
%\changed{Subscripts of $f_0$ removed.}
where $f$ is a coupling strength between the heavy-quark currents and the charmonia 
in the ordinary vacuum without external magnetic fields. 
This strength follows from square of an amplitude 
\begin{eqnarray}
\hspace{-0.8cm}
i \M &=& -  \!\! \int \! \frac{d^4p}{(2\pi)^4} 
\tr \left[ \, \gamma  S_0(p) 
\Gamma (p,p-q)  S_0 (p-q)  \, \right]
\end{eqnarray}
with $\gamma$ and $\Gamma$ meaning 
$i \gamma^5$ and $\Gamma^5$ ($\gamma^\mu$ and $\Gamma^\mu$) 
appearing in a coupling between the pseudoscalar (vector) current and $\eta_c$ ($\Jp$). 
By performing the integrals as in the above computations, we find in both channels 
\begin{eqnarray}
f = \frac{4 m_{c \bar c} N_c }{\pi a_0^3} 
\label{eq:f0}
\ .
\end{eqnarray}

%%%%%%%%%%%%%%%%%%%%%%%%%%%%%%%%%%%%%%%%%%%%%%%%%%%%%%%%%%%%%%%%%%%%%%%

\subsection{Heavy-quark loop}

\label{sec:SE_cal}

We evaluate the self-energy of charmonia in an magnetic field shown in Fig.~\ref{fig:SE}. 
% As in the calculation of the three-point vertex examined in Sec.~\ref{sec:coupling_FS}, 
Diagrammatic calculation is performed in the heavy-quark limit, 
and thus proceeds in a similar way to the calculation in the previous sections.

We shall first examine Diagram (a) in Fig.~\ref{fig:SE}, 
the amplitude of which is written down in Eq.~(\ref{eq:Sigma_a}). 
We carry out an integral with respect to the zeroth component of a loop momentum $p$ 
with a contour enclosed upward. 
% We find that two poles are contained in this path, 
% and that a contribution from a residue at one of the poles is suppressed by 
% a large quark mass compared to the other, as mentioned in Sec.~\ref{sec:coupling_FS}. 
% Thus 
We pick up the leading contribution from a pole located on 
\begin{eqnarray}
\bar p^0 = q^0 - \epsilon_{\bp-\bq}  \simeq m - \eb - \frac{\bp^2}{2m} 
\label{eq:pbar}
\ ,
\end{eqnarray}
providing a residue given by 
\begin{eqnarray}
&& 
\hspace{-0.7cm}
\Res_{p^0 = \bar p^0} \left[ \, \Gamma (p,p-q)  S_0(p-q)  \Gamma ^\dagger (p-q,p) S_2(p) \, \right]
\nonumber
\\
% \hspace{0.5cm}
% &=& \lim_{p^0 \rightarrow \bar p^0} 
% \left[ \, \Gamma (p,p-q) \left( \, (p^0-\bar p^0) S_0(p-q) \, \right)  \Gamma ^\dagger (p-q,p) S_2(p) \, \right]
% \nonumber
% \\
% \hspace{2cm}
&=&
\left[ \, \Gamma (p,p-q)  \left( - P_- \right) \Gamma ^\dagger (p-q,p) 
\left( \lim_{p^0 \rightarrow \bar p^0} S_2(p) \right) \, \right]
\nonumber
\ ,
\\
\end{eqnarray}
% Since $\lim_{p^0 \rightarrow \bar p^0 } (p^2 - m^2) \sim - 2m \left( \eb + \frac{\bp^2}{m} \right)$, 
where we use the projection operators (\ref{eq:prj}) which have properties utilized below, 
$P_\pm P_\mp = 0$, $P_\pm^2 = P_\pm$ and $\gamma^0 P_\pm = P_\pm \gamma^0 = \pm P_\pm$. 
A limiting expression of the quark propagator with two insertions is given by 
\begin{eqnarray}
\lim_{p^0 \rightarrow \bar p^0} S_2(p) 
&\sim&
\frac{1}{4} Q_\EM^2 F_{\alpha \beta} F_{\mu\nu} 
\frac{1}{ \left( \eb + \bp^2 / m  \right) ^5} 
\\
&& \hspace{.3cm} \times
P_+ \left\{ f^\prime_{\alpha\beta \mu\nu} + f^\prime_{\alpha\mu\beta\nu} + f^\prime_{\alpha\mu\nu\beta} \right\} P_+
\nonumber
\end{eqnarray}
with $f^{\prime \, \alpha\beta \mu\nu} = 
\gamma^\alpha P_+ \gamma^\beta P_+ \gamma^\mu P_+ \gamma^\nu$. 
Commuting the gamma matrices as $\gamma^\alpha P_+ = g^{0\alpha} + P_- \gamma^\alpha$, 
we find 
\begin{eqnarray}
P_+ f^{\prime \, \alpha\beta \mu\nu} P_+ 
% &=& P_+ g^{0\alpha}  P_+ \gamma^\beta P_+ \gamma^\mu P_+ \gamma^\nu P_+
= g^{0\alpha} g^{0\beta} g^{0\mu} g^{0\nu} P_+
\ ,
\end{eqnarray}
and thus that the amplitude $\Sigma^{(a)} $ is proportional to 
the vanishing (0,0)-component of the field strength tensors. 
Therefore, the contribution from Diagram (a) vanishes in the leading order in the heavy-quark limit, 
so that we have found in both pseudoscalar and vector states, 
\begin{eqnarray}
\Sigma^{(a)}  \sim 0
\ ,
\end{eqnarray}
and thus $\Sigma \sim \Sigma^{(b)} $.

We shall proceed to examining Diagram (b), the amplitude of which is written down in Eq.~(\ref{eq:Sigma_b}). 
As in the calculation of Diagram (a), we enclose a contour downward 
and pick up a residue at the same pole (\ref{eq:pbar}). 
One would, however, have to compute a residue at a double pole, 
because the quark propagator with an insertion $S_1(p-q)$ has a double-pole structure. 
The residue is thus obtained by operating a derivative as 
\begin{eqnarray}
&&
\hspace{-0.5cm}
\Res_{p^0 = \bar p^0} \, \left[ \, 
\Gamma (p,p-q) S_1(p-q)  \Gamma ^\dagger (p-q,p) S_1(p) \, \right]
\label{eq:double}
\\
% \hspace{2cm}
&=&  
\lim_{p^0 \rightarrow \bar p^0} \, \left[ \, 
\Gamma (p,p-q) \, {\mathcal S} (p-q) \,  \Gamma ^\dagger (p-q,p) \, \frac{ d \ }{ dp^0} S_1(p) \right. 
\nonumber
\\
&& \left. 
\hspace{0.7cm}
+ \Gamma (p,p-q) \left(  \frac{ d \ }{ dp^0}  {\mathcal S} (p-q)  \right)  \Gamma ^\dagger (p-q,p) S_1(p)  
\, \right]
\nonumber
\ ,
\end{eqnarray}
where a shorthand notation is introduced as ${\mathcal S}(p-q) =  (p^0-\bar p^0)^2 S_1(p-q) $. 
% We will examine those two terms one--by--one, 
% and thus divide the amplitude of Diagram (b) into two terms 
% corresponding to the first and second terms in Eq.~(\ref{eq:double}) as 
% \begin{eqnarray}
% - i \Sigma^{(b)} =  -i \Sigma^{(b1)} - i \Sigma^{(b2)} 
% \ .
% \end{eqnarray}
Some ingredients necessary for obtaining the residue follow from 
operation of the limits and derivatives, for the first term in Eq.~(\ref{eq:double}), as 
\begin{eqnarray}
\lim_{p^0 \rightarrow \bar p^0} \frac{ d \ }{ dp^0} S_1(p) 
&=&
- \frac{i}{4} Q_\EM F_{\alpha\beta} \frac{ 1 } { (  2 m  )^2 (\eb+\bp^2/m) ^3}
\label{eq:b11}
\\
&& \hspace{0.3cm} \times
\left[ \,   \left( \eb + \bp^2/m \right)  ( \sigma^{\alpha\beta} \gamma^0 + \gamma^0 \sigma^{\alpha\beta} )
\right. 
\nonumber
\\
&& \left. \hspace{1.2cm} 
+ 4m \left(   \sigma^{\alpha\beta} P_+ + P_+ \sigma^{\alpha\beta} \right) 
\, \right]
\nonumber
\\
\lim_{p^0 \rightarrow \bar p^0 } {\mathcal S} (p-q) &=&
- \frac{i}{4} Q_\EM F_{\alpha\beta}  \frac{ 1 }{   2m   } 
\left\{ \, \sigma^{\alpha\beta}  P_-  +   P_-  \sigma^{\alpha\beta} \, \right\} 
\nonumber
\ ,
\\
\label{eq:b12}
\end{eqnarray}
and, for the second term, as %the limit and derivative are carried out to be, 
\begin{eqnarray}
% \lim_{p^0 \rightarrow \bar p^0} \frac{ d \ }{ dp^0} (p^0-\bar p^0)^2 S_1(p-q) 
&&
\lim_{p^0 \rightarrow \bar p^0} \frac{ d \ }{ dp^0} {\mathcal S} (p-q)
\label{eq:b21}
\\
&& \hspace{.2cm}
% \\
% && \hspace{.3cm} = 
% - \frac{i}{4} Q_\EM F_{\alpha\beta}  \frac{ 1 } { (  2 m  ) ^2}
% \left[ \,  \sigma^{\alpha\beta} ( \gamma^0 + 2 P_- ) + ( \gamma^0 + 2 P_- ) \sigma^{\alpha\beta}  
% \, \right]
=  - \frac{i}{4} Q_\EM F_{\alpha\beta}  \frac{ 1 } { (  2 m  ) ^2}
\left[ \,  \sigma^{\alpha\beta} ( \gamma^0 + 2 P_- ) + ( \gamma^0 + 2 P_- ) \sigma^{\alpha\beta}  
\, \right]
\nonumber
\\
&&
\lim_{p^0 \rightarrow \bar p^0 } S_1(p) \sim
- \frac{i}{4} Q_\EM F_{\alpha\beta}  
\frac{ \left\{ \, \sigma^{\alpha\beta}  P_+  +   P_+  \sigma^{\alpha\beta} \, \right\} }
{   2m \left( \eb + \frac{\bp^2}{m} \right)^2  }
\label{eq:b22}
\ .
\end{eqnarray}

Substituting the limiting behaviors of the propagators (\ref{eq:b11}) and (\ref{eq:b22}) 
into Eq.~(\ref{eq:double}), we find a self-energy of $\eta_c$ as 
\begin{eqnarray}
\label{eq:sig_ps0}
- i \Sigma^{(b)} &=& 
i \frac{Q_\EM^2  }{ 16 m^2} F_{\mu\nu} F_{\alpha\beta}  \, 
\tr\left[ \,  P_-  \sigma^{\mu\nu}   P_- \sigma^{\alpha\beta} \, \right]
\\
&& \hspace{0.3cm} \times 
\int \frac{d^3 \bp}{(2\pi)^3} \vert \psi_\swave (\bp) \vert ^2
\left( \, 2 + \frac{  4m  }{   \left( \eb + \bp^2/m \right) } \, \right) 
\nonumber
\ ,
\end{eqnarray}
and a self-energy of $J/\psi$ as 
\begin{eqnarray}
\label{eq:sig_vec0}
- i \Sigma^{(b) \, \lambda \sigma} 
&=& 
i \frac{Q_\EM^2  }{ 16 m^2} F^{\mu\nu} F^{\alpha\beta} \, \Phi^{\lambda \sigma}_{\mu\nu\alpha\beta} 
% \tr\left[ \, 
% (P_+ \gamma^\lambda P_-)   \sigma^{\mu\nu}  ( P_- \gamma^\sigma  P_+)   \sigma^{\alpha\beta}  
% \, \right]
\\
&& \hspace{0.3cm} \times 
\int \frac{d^3 \bp}{(2\pi)^3} \vert \psi_\swave (\bp) \vert ^2
\left( 2 +   \frac{4m}{\left( \eb + \bp^2/m \right)}  \right) 
\nonumber
\ .
\end{eqnarray}
A trace of the gamma matrices is given by 
$\Phi^{\lambda \sigma}_{\mu\nu\alpha\beta} = \tr\left[ \,(P_+ \gamma^\lambda P_-)   \sigma_{\mu\nu}  
( P_- \gamma^\sigma  P_+)   \sigma_{\alpha\beta}  \, \right]$. 
% where we suppressed the four Lorentz--indices of $\Phi^{\lambda \sigma}$ for simplicity. 
% contracted with the field strength tensors. 
Carrying out the traces, we obtain the self-energies (\ref{eq:sig_ps}) 
and (\ref{eq:sig_vec}) for $\eta_c$ and $J/\psi$, respectively. 
External magnetic fields do not give rise to a self-energy of the unphysical mode of $J/\psi$, 
since the Bethe-Salpeter amplitude vanishes for a temporal mode ($\lambda, \sigma = 0$) when the charmonium is at rest.

We comment on the second-order Stark effect caused by external electric
fields \cite{Pes, BP, LMS92}. This term can be obtained by including the
higher dimensional operator correction to the Bethe-Salpheter equation
in (\ref{eq:G5}) and (\ref{eq:Gvec}) that are proportional to the
external electric field operator and the wave function (see the second
paper in Ref.~\cite{BSamp}).  The effective four point vertex between
the charmonium, external field, charm and the anticharm is given as
\cite{BSamp}
\begin{eqnarray}
M_4^{\nu \mu}=ig \sqrt{\frac{m_{c \bar c}}{N_c}} 
\bigg(\frac{\partial \psi_\swave}{\partial p^\alpha} \bigg) 
F_{\scriptscriptstyle E}^{\nu \alpha } P_+ \Gamma_\mu P_-, 
\end{eqnarray}
where $\Gamma_\mu$ is $\gamma_\mu$ or $i \gamma^5$ for $J/\psi$ or $\eta_c$, respectively. 
Also, the subscript $E$ in the field strength tensor means that only the electric part of 
the field strength tensor is taken.     
Substituting this into the $J/\psi$ self-energy
\begin{eqnarray}
\Sigma & =& 
-i \int \!\! \frac{d^4p}{(2 \pi)^4} {\rm Tr}[M_4^{0\mu} S_0(p-q) M_{4\ \, \mu}^{\dagger 0} S_0(p)] 
\nonumber 
\\
 & =&  -\frac{2m_{\V}}{18} \int_0^\infty \!\!\! d p^2 
\left|\frac{\partial \psi_\swave}{\partial p} \right|^2  \!\!
\frac{p}{\epsilon + p^2/ m}  \left\langle \frac{\alpha}{\pi} E^2 \right\rangle,
\nonumber
\\
\label{eq:Stark}
\end{eqnarray}
which gives the second-order Stark effect formula for the external gauge field \cite{LM09}. 
The same formula is obtained for $\eta_c$. %and spin-projected $J/\psi$ self-energies.

It has been found that the leading-order effect on charmonia by external fields 
is due to external electric field, and that effects of magnetic field are subleading 
in the heavy-quark expansion \cite{LMS92}. 
We find that the self-energies in magnetic fields 
(\ref{eq:sig_ps0})-(\ref{eq:sig_vec0}) 
are also suppressed by a factor of $1/m^2$ compared 
to the second-order Stark effect formula (\ref{eq:Stark}). 
Inserting a field strength tensor of an electric field given by temporal components, 
$F^{0i} = - F^{i0} = E^i$, the trace parts in Eqs.~(\ref{eq:sig_ps0}) and (\ref{eq:sig_vec0}) 
identically vanish when any of the Lorentz indices, $\alpha $, $ \beta $, $ \mu $ or $ \nu$, 
takes temporal component 
because of simple identities $P_\pm \sigma^{0i} P_\pm = \sigma^{0i} P_\mp P_\pm = 0 $, 
showing that there is no additional term contributing to the second-order Stark effect formula.

%%%%%%%%%%%%%%%%%%%%%%%%%%%%%%%%%%%%%%%%%%%%%%%%%%%%%%%%%%%%%%%%%%

%%%%%%%%%%%%%%%%%%%%%%%%%%%%%%%%%%%%%%%%%%%%%%%%%%%%%%%%%%%%%%%%%%

\section{Borel-transformed Wilson coefficients}

\label{sec:W}

In this Appendix, we provide a table of the twist-2 Wilson coefficients $c_n^\ext$ and $c^\ext(\nu)$ 
which are obtained by carrying out the Borel transform of $\tilde \Pi_2$ 
shown in Sec.~\ref{sec:OPEext}. 
% which are respectively obtained from the power correction term to the correlator (see Sec.~\ref{sec:C2}) 
% by taking the derivatives (\ref{eq:Mn}) and the infinite limits (\ref{eq:M0}). 

The moments $c_n^\ext$ are typically represented by the Gauss hypergeometric function ${}_2F_1(a,b,c;\rho)$ 
which is, in conventions in Refs.~\cite{RRY80,RRY81}, defined by 
\begin{eqnarray}
{}_2F_1(a,b,c;\rho) &=& 
\\
&& \hspace{-1.5cm}
\frac{1}{B(a, c-a)} 
\int_0^1 \!\! dt \ t^{a-1} (1-t)^{c-a-1} (1-\rho t)^{-b}
\ \ ,
\nonumber
\end{eqnarray}
where the beta function $B(x,y) $ is related to the gamma function as 
$B(x,y) = \Gamma(x) \cdot \Gamma(y) / \Gamma(x+y) $. 
Hereafter, we suppress the subscripts as $F(a,b,c;\rho) = {}_2F_1(a,b,c;\rho) $ for simplicity. 
Following from the definition of the Borel transform (\ref{eq:Mn}), 
we obtained a useful formula 
\begin{eqnarray}
&&
\frac{ (-1)^n }{ n! } \,  \frac{ d^n \  }{ d \xi^n } 
\left( \ \xi^{-\beta} J_k (\xi) \ \right)
\nonumber
\\
&&  \hspace{0.5cm}
= \frac{ (-1) ^{\beta}  \sqrt{\pi}  }{ 2  } 
\frac{ \Gamma( n + k + \beta ) }{  \Gamma(k) \,  \Gamma ( n + \beta + \frac{3}{2} ) } 
% \label{eq:deri}
\\
&&  \hspace{0.8cm}
 \times ( 1 - \rho )^{  n + 1  } \ 
F  ( n+1 , \frac{3}{2} - k , n + \beta + \frac{3}{2} ; \rho )
\ ,
\nonumber
\end{eqnarray}
for general integers $\beta$ and $k$. 
Relevant variables are 
\begin{eqnarray}
\xi &=& \frac{Q^2}{4m^2} %= \frac{n M^2}{4m^2} = \frac{n}{\nu}
\ \ ,
\\
\rho &=& \frac{\xi}{1+\xi} %= 1- \frac{\nu}{n}
\ \ ,
\\
\nu &=&  \frac{4m^2}{M^2} =  n (1-\rho) 
\ \ ,
\end{eqnarray}
and the Borel mass $M^2 = Q^2 / n$ is maintained being a constant in the infinite limits (\ref{eq:M0}). 
In these limits, the Whittaker function 
\begin{eqnarray}
G(b,c;\nu) &=& \frac{1}{\Gamma(c)} 
\int_0^\infty\!\! e^{-t} t^{c-1} (\nu+t)^{-b} dt
\end{eqnarray}
is related to a limiting behavior of the hypergeometric function as 
\begin{eqnarray}
F(b,\ell,\ell+c;\rho)   \xrightarrow[ n \to \infty]{} \ell^b \, G(b,c;\nu)
\label{eq:FG}
\ \ ,
\end{eqnarray}
so that the Borel-transformed Wilson coefficients $c^\ext(\nu)$ 
can be obtained analytically and represented by the Whittaker function \cite{Bertl}. 
The Wilson coefficients $b_n$ and $b(\nu)$ appearing below 
are shown in Ref.~\cite{RRY81} and Refs.~\cite{Bertl,ML10}, respectively.

% \begin{eqnarray}
% \frac{d \ }{d\nu} G(b,c;\nu) &=&  - b \, G(b+1,c;\nu)
% \\
% G(b,c;\nu)  &\xrightarrow[ \nu \to \infty]{}& \nu ^{-b}
% \end{eqnarray}

\subsection{Pseudoscalar channel (P)}

\begin{eqnarray}
c_n^{\ps , \ext }
% &=& \frac{4}{3} \cdot c_n
% \nonumber
% \\
&=&  \frac{4}{3} \left[ 
b_n - 4n(n+1) (1- \rho) \frac{ F( n+1, - \frac{1}{2}, n+\frac{3}{2}; \rho) }{ F( n, \frac{1}{2}, n + \frac{3}{2}; \rho) }
\right]
\nonumber
\\
\end{eqnarray}

\begin{eqnarray}
c^{\ps, \ext} (\nu) &=&
\frac{ \nu }{2 \, G(\frac{1}{2},\frac{3}{2}; \nu) } 
\left[ \ 
 - G(-\frac{3}{2},\frac{3}{2}; \nu) 
\right. 
\\
&& \left. \hspace{0.25cm}
+ 6 G(-\frac{1}{2},\frac{3}{2}; \nu)  - 8 G(-\frac{1}{2},\frac{1}{2}; \nu) 
\ \right]
\nonumber
\end{eqnarray}

%%%%%%%%%%%%%%%%%%%%%%%%%% Vector

\subsection{Vector channel (V)}

\subsubsection{Longitudinal mode $({\rm V}_\parallel)$}

\begin{eqnarray}
c_n^{\V_{\scriptscriptstyle \! \parallel} , \ext}  &=&
\frac{ 4 n(n+2) (1-\rho) }{ 3 (2n+5) F(n, \frac{1}{2}, n+\frac{5}{2};\rho) }
\\
&&  \hspace{.1cm} \times
\left[ \ 
\frac{ 4 }{ n+2} F(n+1, \frac{1}{2}, n+\frac{5}{2};\rho)
\nonumber
\right.
\\
&& \left. \hspace{.5cm}
- 3 (n+3)(n+4) F(n+1, -\frac{1}{2}, n+\frac{7}{2};\rho)
\right.
\nonumber
\\
&& \left. \hspace{.5cm}
-  (n+3)(n+4) F(n+1, -\frac{3}{2}, n+\frac{7}{2};\rho)
\ \right]
\nonumber
\end{eqnarray}

\begin{eqnarray}
c^{\V_{\scriptscriptstyle \!  \parallel }, \ext}  (\nu) &=& 
\frac{ 2 \nu }{ 3 \, G(\frac{1}{2}, \frac{5}{2} ; \nu) } \left[ \ 
8 \, G(\frac{1}{2}, \frac{3}{2} ; \nu) 
\right.
\\
&& \left. \hspace{.5cm}
- 3 \,  G( - \frac{1}{2}, \frac{5}{2} ; \nu) 
-  G( - \frac{3}{2}, \frac{5}{2} ; \nu) 
\ \right]
\nonumber
\end{eqnarray}

\subsubsection{Transverse mode $({\rm V}_\perp)$}

\begin{eqnarray}
c_n^{\V_{\scriptscriptstyle \! \perp }, \ext}  &=& 
\frac{ 4 n (n+2) (1-\rho) }{ 3 (2n+5) F(n, \frac{1}{2}, n+\frac{5}{2};\rho) }
\\ 
&& \hspace{.2cm} \times
\left[ \ 
6   F(n+1, \frac{1}{2}, n+\frac{7}{2};\rho) 
\nonumber
\right.
\\
&& \left. \hspace{1.cm}
+6 (n+3) F(n+1, -\frac{1}{2}, n+\frac{7}{2};\rho)
\right.
\nonumber
\\
&& \left. \hspace{1.cm}
-  (n+3)(n+4) F(n+1, -\frac{3}{2}, n+\frac{7}{2};\rho)
\ \right]
\nonumber
\end{eqnarray}

\begin{eqnarray}
c^{\V_{\scriptscriptstyle \! \perp} , \ext}  (\nu) &=&
\frac{ 2 \nu }{ 3 \, G(\frac{1}{2}, \frac{5}{2} ; \nu) } \left[ \ 
6 \, G(\frac{1}{2}, \frac{5}{2} ; \nu) 
\right.
\\
&& \left. \hspace{.5cm}
+ 6 \,  G( - \frac{1}{2}, \frac{5}{2} ; \nu) 
-  G( - \frac{3}{2}, \frac{5}{2} ; \nu) 
\ \right]
\nonumber
\end{eqnarray}

%%%%%%%%%%%%%%%%%%%%%%%%%%%%%%%%%%%%%%%%%%%%%%%%%%%%%%%%%%%%%%%%%%

%%%%%%%%%%%%%%%%%%%%%%%%%%%%%%%%%%%%%%%%%%%%%%%%%%%%%%%%%%%%%%%%%%%%%%%%%%%%%%%%

%%%%%%%%%%%%%%%%%%%%%%%%%%%%%%%%%%%%%%%%%%%%%%%%%%%%%%%%%%
%%%%%%%%%%%%%%%%%%%%%%%%%%%%%%%%%%%%%%%%%%%%%%%%%%%%%%%%%%

%bib

\input{bib}
\end{document}

%% file: head.tex
\vspace*{-10mm}
\begin{flushright}
%KEK-TH-1637
\end{flushright}
\vspace{-5mm}

%%%%%%%%%%%%%%%%%%%%%%%%%%%%%%%%%%%%%%%%%

\author{Sungtae Cho} \email{\tt
sungtae.cho@kangwon.ac.kr}

\affiliation{ Institute of Physics and Applied Physics, Yonsei
University, Seoul 120-749, Korea } \affiliation{ Division of
Science Education, Kangwon National University, Chuncheon 200-701,
Korea }

\author{Koichi Hattori} \email{\tt koichi.hattori@riken.jp}

\affiliation{
Institute of Physics and Applied Physics,
Yonsei University, Seoul 120-749, Korea
}
\affiliation{
Theoretical Research Division, Nishina Center, RIKEN, Wako, Saitama 351-0198, Japan
and
RIKEN BNL Research Center, Bldg. 510A, Brookhaven National Laboratory, Upton, New York 11973, USA
}

%%%%%%%%%%%%%%%%%%%%%%%%%%%%%%%%%%%%%%%%%

\author{Su Houng Lee} \email{\tt suhoung@yonsei.ac.kr}

\affiliation{
Institute of Physics and Applied Physics,
Yonsei University, Seoul 120-749, Korea
}

%%%%%%%%%%%%%%%%%%%%%%%%%%%%%%%%%%%%%%%%%

\author{Kenji Morita} \email{\tt kmorita@yukawa.kyoto-u.ac.jp}

%\affiliation{
%Yukawa Institute for Theoretical Physics, Kyoto University, Kyoto 606-8502, Japan
%}
\affiliation{Frankfurt Institute for Advanced
Studies, Ruth-Moufang-Str. 1, D-60438 Frankfurt am Main, Germany}
\affiliation{
Institute of Theoretical Physics, University of Wroclaw,
PL-50204 Wroclaw, Poland
}
\affiliation{
Yukawa Institute for Theoretical Physics, Kyoto University, Kyoto 606-8502, Japan
}

\author{Sho Ozaki} \email{\tt sho@post.kek.jp}
\affiliation{
Institute of Physics and Applied Physics,
Yonsei University, Seoul 120-749, Korea
}
\affiliation{
Theory Center, IPNS, High energy accelerator research organization (KEK),
1-1 Oho, Tsukuba, Ibaraki 305-0801, Japan
}

%%%%%%%%%%%%%%%%%%%%%%%%%%%%%%%%%%%%%%%%%

%+++++++++++++++++++++++++++++++++++++++++++++++++++++++%

\vspace*{5mm}
\title{Charmonium Spectroscopy in Strong Magnetic Fields
\\by QCD Sum Rules: S-Wave Ground States}

%+++++++++++++++++++++++++++++++++++++++++++++++++++++++%

\date{\today}% It is always \today, today,
 %  but any date may be explicitly specified

\begin{abstract}
We investigate quarkonium mass spectra in external constant magnetic fields by using QCD sum rules.
We first discuss a general framework of QCD sum rules necessary for %investigating any
properly extracting meson spectra from current correlators
computed in the presence of strong magnetic fields,
that is, a consistent treatment of mixing effects caused in the mesonic degrees of freedom.
% in external magnetic fields on the phenomenological side.
We then implement operator product expansions for pseudoscalar and vector
heavy-quark current correlators by taking into account external constant magnetic fields as operators,
and obtain mass shifts of the lowest-lying bound states $\eta_c$ and $J/\psi$
in the static limit with their vanishing spatial momenta.
Comparing results from QCD sum rules with those from hadronic effective theories,
we find that the dominant origin of mass shifts comes from a mixing between $\eta_c$
and $J/\psi$ with a longitudinal spin polarization, accompanied by other subdominant effects
such as mixing with higher excited states and continua.
\end{abstract}

%+++++++++++++++++++++++++++++++++++++++++++++++++++++++%

%\preprint{}

%\pacs{}
% PACS, the Physics and Astronomy
% Classification Scheme.
%\keywords{Strong magnetic field, Vacuum birefringence, Landau levels}
%Use showkeys class option if keyword display desired

%%%%%%%%%%%%%%%%%%%%%%%%%%%%%%%%%%%%%%%%%%%%%%%%%%%%%%%%%%%%

\maketitle